\documentclass[]{aastex631}

\usepackage{multirow}

\shorttitle{Exploring the recycling model}
\shortauthors{Madeira et al.}

\defcitealias{hesselbrock2017}{HM17}
\graphicspath{{./}{figures/}}

\begin{document}

\title{Exploring the recycling model of Phobos formation: rubble-pile satellites
\footnote{Released on \today}}

\author[0000-0001-5138-230X]{Gustavo Madeira}
\affiliation{Universit\'e de Paris, Institut de Physique du Globe de Paris, CNRS
F-75005 Paris, France}
\affiliation{Grupo de Din\^amica Orbital e Planetologia, S\~ao Paulo State University (UNESP), 333 Av. Dr. Ariberto Pereira da Cunha, Guaratinguet\'a, SP 12516-410, Brazil}

\author[0000-0002-7442-491X]{S\'ebastien Charnoz}
\affiliation{Universit\'e de Paris, Institut de Physique du Globe de Paris, CNRS
F-75005 Paris, France}

\author[0000-0003-4045-9046]{Yun Zhang}
\affiliation{Department of Aerospace Engineering, University of Maryland, College Park, MD 20742, USA}
\affiliation{Universit\'e C\^ote d`Azur, Observatoire de la C\^ote d`Azur, CNRS, Laboratoire Lagrange, Nice, 06304, France}

\author[0000-0003-4590-0988]{Ryuki Hyodo}
\affiliation{ISAS/JAXA, Sagamihara, Kanagawa, Japan}

\author[0000-0002-0884-1993]{Patrick Michel}
\affiliation{Universit\'e C\^ote d`Azur, Observatoire de la C\^ote d`Azur, CNRS, Laboratoire Lagrange, Nice, 06304, France}

\author[0000-0001-6702-0872]{Hidenori Genda}
\affiliation{Earth-Life Science Institute, Tokyo Institute of Technology, Meguro-ku, Tokyo 152-8550, Japan}

\author[0000-0002-3949-6045]{Silvia Giuliatti Winter}
\affiliation{Grupo de Din\^amica Orbital e Planetologia, S\~ao Paulo State University (UNESP), 333 Av. Dr. Ariberto Pereira da Cunha, Guaratinguet\'a, SP 12516-410, Brazil}

\begin{abstract}
Phobos is the target of the return sample mission Martian Moons eXploration by JAXA that will analyze in great details the physical and compositional properties of the satellite from orbit, from the surface and in terrestrial laboratories, giving clues about its formation. Some models propose that Phobos and Deimos were formed after a giant impact giving rise to an extended debris disk. Assuming that Phobos formed from a cascade of disruptions and re-accretions of several parent bodies in this disk, and that they are all characterized by a low material cohesion, \cite{hesselbrock2017} have showed that a recycling process may happen during the assembling of Phobos, by which Phobos' parents are destroyed into a Roche-interior ring and reaccreted several times. In the current paper we explore in details the recycling model, and pay particular attention to the characteristics of the disk using 1D models of disk/satellite interactions. In agreement with previous studies we confirm that, if Phobos' parents bodies are gravitational aggregates (rubble piles), then the recycling process does occur. However, Phobos should be accompanied today by a Roche-interior ring. Furthermore, the characteristics of the ring are not reconcilable with today`s observations of Mars' environment, which put stringent constraints on the existence of a ring around Mars. The recycling mechanism may or may not have occurred at the Roche limit for an old moon population, depending on their internal cohesion. However, the Phobos we see today cannot be the outcome of such a recycling process.
\end{abstract}

\keywords{planets and satellites: formation -- planets and satellites: dynamical evolution and stability -- planets and satellites: individual (Phobos, Deimos, Mars) -- planets and satellites: rings -- planets and satellites: terrestrial planets}

\section{Introduction} \label{sec:intro}
Discovered in 1878 \citep{hall1878}, Phobos and Deimos are the two natural satellites of Mars and target bodies of the return sample mission Martian Moons eXploration (MMX), from the Japan Aerospace eXploration Agency \citep[JAXA,][]{kuramoto2022}. The satellites are relatively small with low bulk density (probably high porosity), irregular shape, and cratered surface. Phobos' radius is $\sim$11~km, while Deimos has a radius of $\sim$6~km \citep{thomas1993,willner2014}. The latter is located at about $6.92$ Mars' radius \citep[${\rm R_M}=3389.5$~km,][]{Peale1999,Seidelmann2002}, presenting a weak tidal interaction with the planet. Because of this, the satellite moves very slowly away from Mars. Phobos, in turn, is close to the planet \citep[$\sim2.77~{\rm R_M}$,][]{Peale1999}, felling a tidal torque responsible for its orbital decay. The difference in direction of migration of the satellites is because Phobos resides inside Mars' synchronous orbit (at about $6.03~{\rm R_M}$) while Deimos resides beyond it. Tidal evolution studies show that Phobos would fall to Mars in less than 40~Myr \citep{sharpless1945,Samuel2019,bagheri2021}. The most likely scenario, however, is the destruction of Phobos by tides before reaching Mars surface, between $1.0~{\rm R_M}$ and $2.0~{\rm R_M}$. A significant fraction of the satellite is believed to be heavily damaged \citep{black2015} and, as a consequence, Phobos must either be fully destroyed in a cloud of particles or fragmented into several large fragment accompanied by a cluster of debris \citep{black2015}. This leaves the question: Are we in a privileged time, in which we can observe Phobos just before its end?

Early models \citep{pang1978,burns1992,pajola2012,pajola2013} proposed that Phobos and Deimos were asteroids captured by Mars. This hypothesis was motivated by data obtained by spacecraft, such as Mariner 9, Viking 1 and 2, and Phobos 2, which show that martian moons has low albedo and spectra resembling carbonaceous asteroids \citep{fraeman2012,witasse2014}. However, the capture model is not consistent with dynamical constraints, as tidal dissipation is not strong enough to change Deimos' orbit from highly inclined and eccentric -- as expected for a captured object -- to near-equatorial and near-circular in a shorter time than the Solar System age \citep{szeto1983}. Some modified mechanisms have been proposed, such as a three-body capture \citep{hansen2018}. However, such a capture remains unlikely; in particular, it is unclear whether Phobos and Deimos would survive a capture by Mars if they have a rubble pile structure \citep{zhang2020}.

In opposition to the capture mechanism, it has been proposed that Phobos and Deimos formed in a debris disk around Mars, which would naturally explain their near-circular and near-equatorial orbits \citep{craddock1994}. \cite{singer2003} proposed that the debris disk around Mars would originate from an object captured and destroyed at the Mars' Roche limit due to tidal effects. The recent work of \cite{bagheri2021} propose that Phobos and Deimos originated from the disruption of a progenitor moon, likely formed \textit{in situ} around Mars. By performing backward tide simulations, they obtain that the orbits of Phobos and Deimos convert to a common position above the synchronous orbit in $\sim$1-3~Gyr, the possible time of destruction of the progenitor moon. After the destruction, Phobos and Deimos would be launched in highly eccentric orbits, feeling a strong satellite tidal dissipation, responsible for the damping of eccentricity and generating an inward migration. When the orbits are sufficiently circularized, the Martian tidal dissipation exceeds that of the satellite, and Deimos starts to migrate outward, while Phobos, now below the synchronous orbit, migrates inward toward its actual position. However, the post-evolution of Phobos and Deimos after the split was studied in detail by \cite{hyodo2022}, who obtained that the satellites collide with each other in less than $10^4$~yr in most of the cases analysed, forming a debris ring around the synchronous radius. Therefore, Phobos and Deimos are unlikely to be the direct result of the splitting of an ancestor moon \citep{hyodo2022}.

As a matter of fact, the large basins seen on the surface of Mars seem to indicate past highly energetic impacts between the planet and ongoing objects. In particular, the impact responsible for the Borealis basin, which covers almost 40\% of the Mars' surface \citep[hemispheric crustal dichotomy,][]{marinova2008}, would provide enough energy for rock vaporization, resulting in the formation of a debris disk \citep{craddock2011}. Furthermore, Mars' current spin rate can only be explained by an impact with an external object \citep{dones1993,craddock2011}, placing a giant impact as the only mechanism capable of offering the appropriate environment for the formation of Phobos and Deimos.

Several works \citep{citron2015,rosenblatt2016,hyodo2017,canup2018} have used impacts models relying on the Smoothed Particle Hydrodynamics (SPH) techniques to recreate the impact responsible for the Borealis basin and Mars' current spin rate. The ejecta produced by the impact -- composed of material from Mars and the impactor -- would reside right after the impact in highly elliptical and inclined orbits around the planet. Ejecta in the form of molten droplets due to their high temperature \citep[$\sim$2000~K,  ][]{hyodo2017,hyodo2018}, then begin to solidify and collide with each other, inducing energy loss while angular momentum is conserved. As a consequence, the eccentricities and inclinations are quickly dampened, resulting in the formation of a flat and extended disk of debris with mass $\sim$10$^{20}$~kg. Whereas most of the disk's mass is located inside the Roche limit of Mars, some debris from the impact extends up to the synchronous orbit \citep{citron2015,rosenblatt2016,hyodo2017}.

The debris disk is expected to viscously spread due to inter-particle collisions, and a mechanism similar to that proposed for the formation of Saturn's icy moons \citep{charnoz2010,charnoz2011} can be assumed to take place at the Martian Roche limit, located at about $3.2~{\rm R_M}$. The ring located inside the Roche limit spreads viscously, releasing material outside the Roche limit. This material coagulates due to gravitational instabilities, generating the accretion of small-sized moonlets. In the case of Saturn -- Roche limit outside the synchronous orbit -- moonlets grow by mutual collisions and migrate outward due to disk-satellite torques and tidal effects, giving rise to the icy moons \citep{charnoz2010,charnoz2011}. However, in the case of Mars, the Roche limit lies within the synchronous orbit, and the moonlets feel opposing forces: the disk-satellite torques push them outside the planet and tidal effects cause an inward migration of moonlets located within the synchronous orbit. As a consequence, there is a maximum distance at which a moon can migrate beyond the Roche limit, well inside the current orbit of Deimos. A satellite can be driven to Deimos' position only by resonant trains, in which inner moons slowly migrating outwards capture the outermost satellite into mean motion resonances (MMRs hereafter), pushing it towards outside the synchronous radius \citep{Salmon2017}. Forming Deimos in such a scenario shows to be a trammel \citep{rosenblatt2012}, which spurred the development of more sophisticated mechanisms for the formation of Phobos and Deimos.

The stirred debris disk model proposed by \cite{rosenblatt2016} assumes an initial outer disk -- portion of the disk beyond the Roche limit -- with a mass greater than that obtained by \cite{citron2015}. The outer disk is assumed to be composed of a population of embryos located from $4.2~{\rm R_M}$ to $7~{\rm R_M}$. Due to viscous spreading and gravitational instabilities, moonlets are formed just outside the Roche limit. Eventually, all the moonlets result in a single massive moon that migrates outward due to disk-satellite torques. Embryos are trapped in the 2:1 and 3:2 MMR with the massive moon and migrate outward with it. Material is collected and accreted inside the massive moon's MMR.  At some point, the system is composed only by a disk, massive moon and two moons formed near the 2:1 and 3:2 MMR with the massive moon. When the disk is sufficiently depleted, the tidal torque exceeds the disk-satellite torques and the massive moon migrates inward, pushing the disk towards and eventually falling onto the planet, leaving only the two satellites in the system. In the original \citeauthor{rosenblatt2016} paper, the moons are assumed to be very cohesive, so they do not disrupt when they cross Mars' Roche Limit, and fall down to the planet's surface. This hypothesis is opposite to the one of the \cite{hesselbrock2017} -- \citetalias{hesselbrock2017} hereafter-- that assume low-cohesion moons, and thus result in a ``recycling'' process (see below).
In $\sim 1.4$\% of their simulations, \cite{rosenblatt2016} find innermost and outermost satellites with masses lying within 5\% and 30\% the current ones, respectively. The obtained location are equal to those expected for the satellites in the past.

The Roche limit is a theoretical distance at which a fluid body no longer has a tidal equilibrium shape, which means that the fluid object will be destroyed by tidal effects upon reaching this distance. For this reason, the Roche limit is also called the ``Fluid Roche limit'' (FRL hereafter), as it will be called from now on in this article. Solid bodies are stronger than fluids and can be destroyed at smaller distances \citep{holsapple2006,holsapple2008}, at the ``Rigid Roche limit'' (RRL hereafter). Using this distinction between FRL and RRL, \citetalias{hesselbrock2017} propose another model for Phobos formation, the "recycling model". According to them, Deimos is a direct fragment from the giant impact, while Phobos formed from the debris disk initially confined within the FRL.

Through the aforementioned mechanism, moonlets form outside the FRL and collide with each other, giving rise to a moon. This moon eventually begins to migrate inward, reaching the RRL. Assuming that the moon has low-cohesion, the moon breaks and forms a new ring that spreads, restarting the full process. For each destruction/accretion cycle, the total mass of the system is divided by $\approx 5$ in general. This factor is a direct consequence of how the viscous effects are handled, being susceptible to the system parameters. The mass of the largest moon in the oldest cycle depends on the initial width of the ring \citep{hyodo2015}. The system evolution will also depend on the RRL location. It is expected that the closer to the planet the satellite is destroyed, more material will be deposited on Mars. \citeauthor{hesselbrock2017} obtain in their representative simulation (RRL=$1.6~{\rm R_M}$) that six of these recycling cycles are needed to form the current Phobos, meaning that there were five Phobos ancestors in the past. Now, if the RRL position is assumed to be $1.2~{\rm R_M}$, e.g., they find that for each cycle, the mass of the system is divided by $\approx 17$, taking 3 cycles to form Phobos.

An important aspect of the recycling model is the fact that \citetalias{hesselbrock2017} obtain a ring coexisting with Phobos with an optical depth $\tau\sim0.03$. Observations of Mars' environment show that no ring is detectable around Mars with optical depth $\tau>3\times 10^{-5}$ \citep{duxbury1988} and no particle is detected around Mars down to a detection limit of 75~m \citep{showalter2006}.  Therefore, the ring obtained by \citetalias{hesselbrock2017} would be detectable, which can be assumed to be a strong enough caveat to rule out the model.
In the original \citetalias{hesselbrock2017} paper, only one disk case was investigated, and due to computer limitation the full cycle could not be computed over 4.5 Gyr evolution. The model from \citetalias{hesselbrock2017} is mostly the same as presented in \cite{salmon2010} and \cite{charnoz2010}, but using a different version of the code. In this work, we revisit the recycling model, searching for a good set of parameters capable of forming Phobos and a non-visible ring. We emphasize that we use our original \texttt{HYDRORINGS} code \citep{salmon2010,charnoz2010}, which is able to compute the ring's evolution over the full history of the Solar System.

In this work, we focus on the recycling model, analyzing the evolution of the satellites and studying the properties of the residual ring obtained by the process. The existence of different cycles relies on the important assumption that satellites are destroyed by tidal forces before falling onto the planet. This assumption distinguishes the recycling model from the stirred debris disk model. If we assume that the massive moon in the stirred debris disk is also destroyed by tidal forces, the formation of a ring of debris occurs and, consequently, material recycles occur as well.

What defines whether an object will fall entirely onto Mars or not is its constitutive characteristics. Weakly cohesive objects are expected to be disrupted, while strongly cohesive objects can survive \citep{black2015}. MMX plans to collect and return samples from Phobos surface, which will allow constraining the satellite's strength \citep{hyodo2019,hyodo2021} and obtaining evidence supporting one of the models for Phobos' formation. Although this is never mentioned by \cite{rosenblatt2016} and \citetalias{hesselbrock2017}, they are assuming different strengths for objects accreted in the debris disk. Here, we explore the case of rubble-pile satellites.

The structure of the paper is as follows: In Section~\ref{sec:rrl}, we explain how we obtain the disruption distance depending on the mass and friction angle of a rubble-pile. In Section~\ref{sec:disruption}, we analyze the tidal evolution of Phobos and explore a tidal downsizing process of the satellite. In Section~\ref{sec:pkdgrav} we check the analytical calculations by running simulations using the gravitational N-body code \texttt{pkdgrav} and its implementation of the Soft-Sphere Discrete Element Method to model the evolution of rubble piles \citep{Richardson2000,Schwartz2012,Zhang2017,Zhang2018} and analyze the Phobos downsizing. In Section~\ref{sec:recycling} we simulate the full recycling mechanism using the \texttt{HYDRORINGS} code \citep{salmon2010,charnoz2010} and vary disk mass, debris size and friction angle. We detail the rings evolution with time in different scenarios. In Section~\ref{sec:discussion}, we discuss the implications of our results and perform analysis on the Yarkovsky effect and the effects of the recycling process on Deimos. We present our conclusions in Section~\ref{sec:conclusion}.

\section{Disruption location of a rubble-pile satellite} \label{sec:rrl}

In this section we compute the distance from Mars, at which a rubble-pile satellite would be disrupted depending on its material properties and shape.

Close to the Roche limit, the planet's tidal forces tend to stretch satellites, making their equilibrium shape non-spherical. Effects such as rotation and tidal forces induce the redistribution of the satellite material. In the hypothetical case of a fluid satellite, planet's tidal forces deform the satellite until it reaches an equilibrium shape: a Roche or Jean ellipsoid \citep{chandrasekhar1969}. Satellites very close to the planet are not able to reach an equilibrium shape and are torn apart by tidal torques. The distance, from the planet's center, at which a fluid satellite no longer has an equilibrium shape is the "Fluid Roche Limit" (FRL).

The fluid model is mostly relevant for large objects, typically $>100$~km, whose cohesion is negligible compared to their self-gravity. Conversely, small objects may be partially or mostly sustained by their internal strength. In solid bodies, strength is defined as the object's ability to withstand stresses. There are three strengths that define the total strength of an object: the tensile strength, cohesion (the shear strength at zero pressure), and compressive strength, see e.g. \cite{dobrovolskis1990,sridhar1992,holsapple2006}. In this way, solid satellites can withstand tidal forces at inner distances to the FRL.

Aiming to analyze whether a satellite can survive a certain distance without being torn apart, we start by evaluating the forces on the object: centrifugal force due to spin, self-gravity, and tidal force from the planet. We assume that the satellite has an ellipsoidal shape (with semi-axes $a_1\geq a_2\geq a_3$), mass-to-planet ratio $p$, and uniform bulk density $\rho$. The satellite is in a circular and equatorial orbit in the $x_1x_2$-plane, where $d$ is the planet-satellite distance. It is also spinning around the $x_3$ axis with synchronous rotation.

The force on the body, per unit of mass, is \citep{holsapple2006}:
\begin{equation}
b_i=\left[-2\pi\rho GA_i+\frac{GM}{d^3}S_i\right]x_i
\end{equation}
where $G$ is the gravitational constant, $M$ is the planet mass, and the coefficients $S_i$ are: $S_1=3+p$, $S_2=p$, and $S_3=-1$. The self-gravity coefficients $A_i$ are given by \citep{holsapple2006}:
\begin{equation}
A_i=a_1a_2a_3\int_0^\infty\frac{du}{(a_i^2+u)(a_1^2+u)^{1/2}(a_2^2+u)^{1/2}(a_3^2+u)^{1/2}}
\end{equation}

The equilibrium equation, relating the stress tensor components $\sigma_{ij}$ and the force on the body \citep{holsapple2001} is :
\begin{equation}
\frac{\partial}{\partial x_j}\sigma_{ij}+\rho b_i=0   \label{tensor}
\end{equation}

Using Equation~\ref{tensor} and averaging over the body volume, we find that \citep{holsapple2008}
\begin{equation}
\overline{\sigma}_{ii}=\left[-2\pi\rho GA_i+\frac{GM}{d^3}S_i\right]\frac{\rho a_i^2}{5}
\end{equation}
while the non-diagonal components are equal to zero.

Once we know the average stress on the satellite ($\overline{\sigma}_{ij}$), we can compare it to the satellite  strength. For this, we assume the Drucker-Prager criterion, a failure criterion for geological materials given by \citep{chen2007}
\begin{equation}
\sqrt{(\overline{\sigma}_{xx}-\overline{\sigma}_{yy})^2+(\overline{\sigma}_{yy}-\overline{\sigma}_{zz})^2+(\overline{\sigma}_{zz}-\overline{\sigma}_{xx})^2}\leq \sqrt{6}[k_{dp}-s_{dp}(\overline{\sigma}_{xx}+\overline{\sigma}_{yy}+\overline{\sigma}_{zz})]
\end{equation}
where $s_{dp}$ and $k_{dp}$ are material constants related to cohesion $Y$ and friction angle $\phi$. These two parameters are related to inter-particle forces within the body, with cohesion giving the body's response to shear stress at zero normal stress, being related to inter-molecular forces. The friction angle is responsible for measuring the response of the body under shear stress and is related to the geometrical interlocking of the granular particles \citep{Sanchez2016}. The constants $s_{dp}$ and $k_{dp}$ are given by \citep{chen2007}
\begin{equation}
s_{dp}=\frac{2\sin\phi}{\sqrt{3}(3-\sin\phi)}
\end{equation}
and
\begin{equation}
k_{dp}=\frac{6Y\cos\phi}{\sqrt{3}(3-\sin\phi)}
\end{equation}

In our numerical simulations, we assume the typical cohesion of a rubble-pile object, $Y=0.025\rho^2/(a_1a_2a_3)^{1/6}$ \citep{holsapple2008,black2015}, and set three values of friction angle: $\phi=25^{\circ}$, approximately the friction angle of a close-packed rubble-pile composed by frictionless particles \citep{albert1997}, $\phi=40^{\circ}$, a typical value for rocks \citep{usoltseva2019}, and $\phi=80^{\circ}$, corresponding to a hypothetical case of an extremely packed core since $\phi\gtrsim60^{\circ}$ are not generally found in nature \citep{yang2018,usoltseva2019}.

\begin{figure}[ht!]
\gridline{\fig{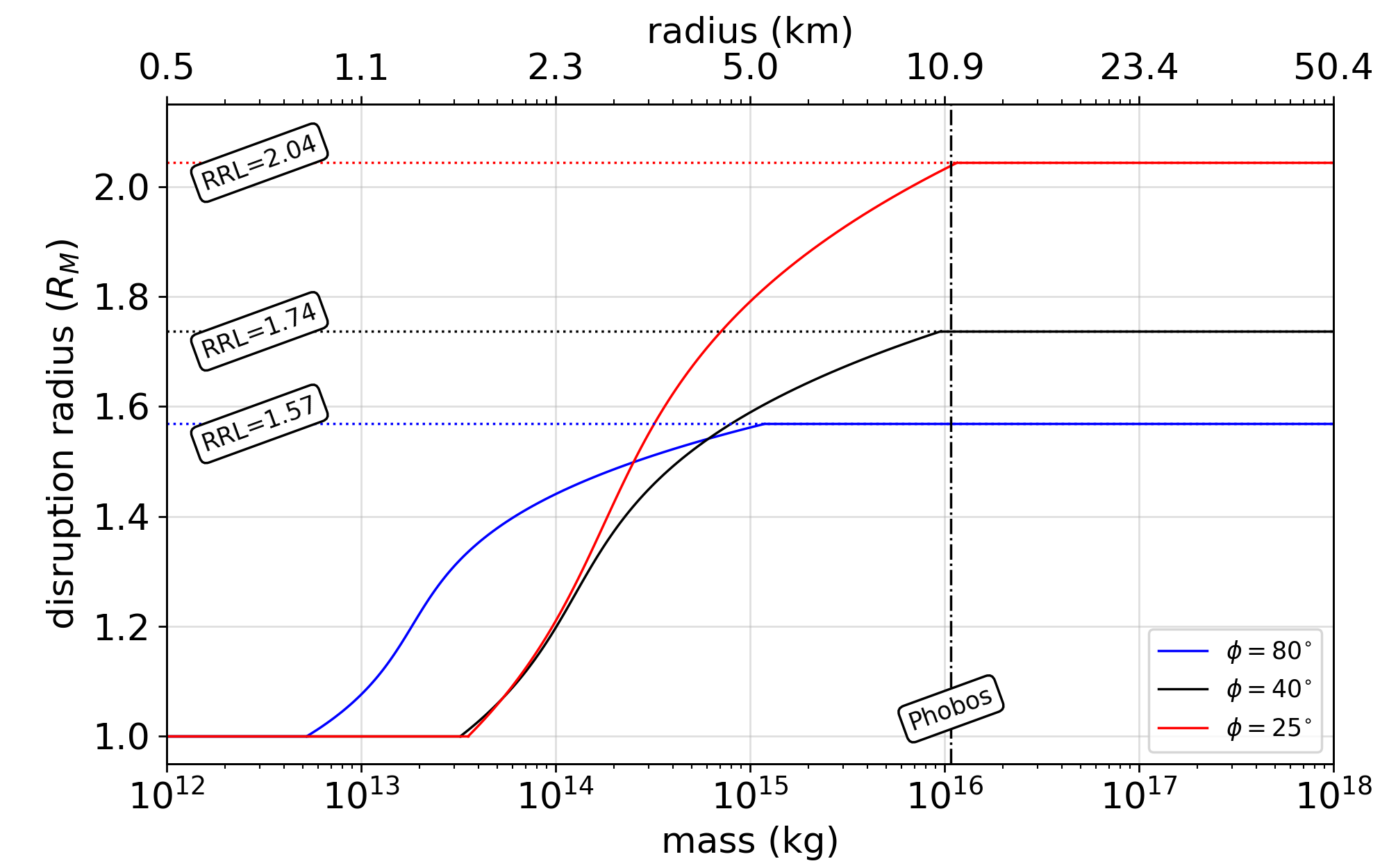}{0.8\textwidth}{}}
\caption{Disruption location, in Mars radius (${\rm R_M}$), of a rubble-pile satellite as a function of its mass/equivalent spherical radius. The red, black, and blue lines correspond to the cases with friction angle $\phi=25^{\circ}$, $40^{\circ}$, and $80^{\circ}$, respectively. The dot-dashed vertical line shows the mass and radius of Phobos. \label{fig:rrl}}
\end{figure}

We look for the disruption location of a rubble-pile object around Mars by using the following methodology: For a given mass, we find all possible combinations of semi-axes for the shape of the object. For this, we assume $a_2$ and $a_3$ in the ranges $0<a_2/a_1<1$ ($\Delta(a_2/a_1)=0.01$) and $0<a_3/a_1\leq a_2/a_1$ ($\Delta(a_3/a_1)=0.01$), with $a_1$ calculated from the mass. Then we vary the semi-major axis of the object from $2.2~{\rm R_M}$ to $1.0~{\rm R_M}$, with a step of $0.05~{\rm R_M}$ (distances are counted from external to internal ones), and apply the Drucker-Prager criterion for all configurations. With this, we obtain which ellipsoidal shapes are stable at that location. Then, we assume as ``Rigid Roche Limit'' the distance below which none of the configurations is no longer stable. We do this calculation for 400 different values of mass, equally spaced in the range $10^{12}-10^{18}$~kg, assuming the bulk density of Phobos $\rho=1.845~{\rm g/cm^3}$ \citep{willner2014,dmitrovskii2022}. Figure~\ref{fig:rrl} shows the disruption distance as a function of satellite's mass/equivalent spherical radius for $\phi=25^{\circ}$ (red line), $40^{\circ}$ (black line), and $80^{\circ}$ (blue line). The vertical line is set at the mass/radius of Phobos.

For each value of the friction angle, there is a critical body mass beyond which the RRL is constant (and thus mass independent, but depending on $\phi$), and below which the RRL is an increasing function of the body mass. All solid bodies contain a distribution of incipient flaws and the size of the largest flaw increases with the object's size. Since the weakness of a body is defined by the size of the flaws \citep{holsapple2008}, we have that large bodies are weaker than small ones. Small bodies behave in the ``strength regime'' and increasing in size, the strength decreases, and there is a threshold above which it doesn't matter and gravity takes the lead. This corresponds to the ``gravity regime'' in which the greater the friction angle, the closer to the planet the satellite disrupts, as can be seen in Fig.~\ref{fig:rrl}.

Therefore, RRL=$2.04~{\rm R_M}$, $1.74~{\rm R_M}$, and $1.57~{\rm R_M}$ for $\phi=25^{\circ}$, $40^{\circ}$, and $80^{\circ}$, respectively. \citetalias{hesselbrock2017} set RRL=$1.6~{\rm R_M}$, approaching our hypothetical case with $\phi=80^{\circ}$. In the numerical simulations of Section~\ref{sec:recycling}, we assume that a satellite breaks up at the location given by Figure~\ref{fig:rrl}, which means that satellites in the strength regime will disrupt in different locations than those in the gravity regime.

\section{Exploration of the tidal decay and erosion of one single moon} \label{sec:disruption}
In the previous section we computed the distance at which a rubble pile would be disrupted by tidal forces. We now focus on the coupling of the inward tidal migration of the moon with the change of its shape, as it falls down to the planet. Orbital tidal decay is dynamically a very slow process. For example, it will take about 15~Myr for Phobos to move from $2.5~{\rm R_M}$ to $2~{\rm R_M}$, that is about 19 billions of orbits \citep{black2015}. As the object deforms, it may have time to re-organize its shape to the evolving tidal stress environment. So, we may expect that tidal forces result in a slow erosion of the object (a diminution of its average radius), as its upper layers are slowly tidally eroded. We call this process ``tidal downsizing'', being similar to the tidal stripping process obtained by \cite{Canup2010} for differentiated satellites. According to \citeauthor{Canup2010}'s work, tides are responsible for stripping material from the outer layers of satellites within the Roche limit, since these layers have a lower density than the rest of the body. Previous studies \citep{black2015} have shown that Phobos would reach the planet's surface in about 30 Myr. However the mass and radius of Phobos were considered as a constant in this study. Here, we re-assess Phobos' evolution, taking into consideration the progressive tidal downsizing of the object below the Roche Limit.

\begin{figure}[ht!]
\gridline{\fig{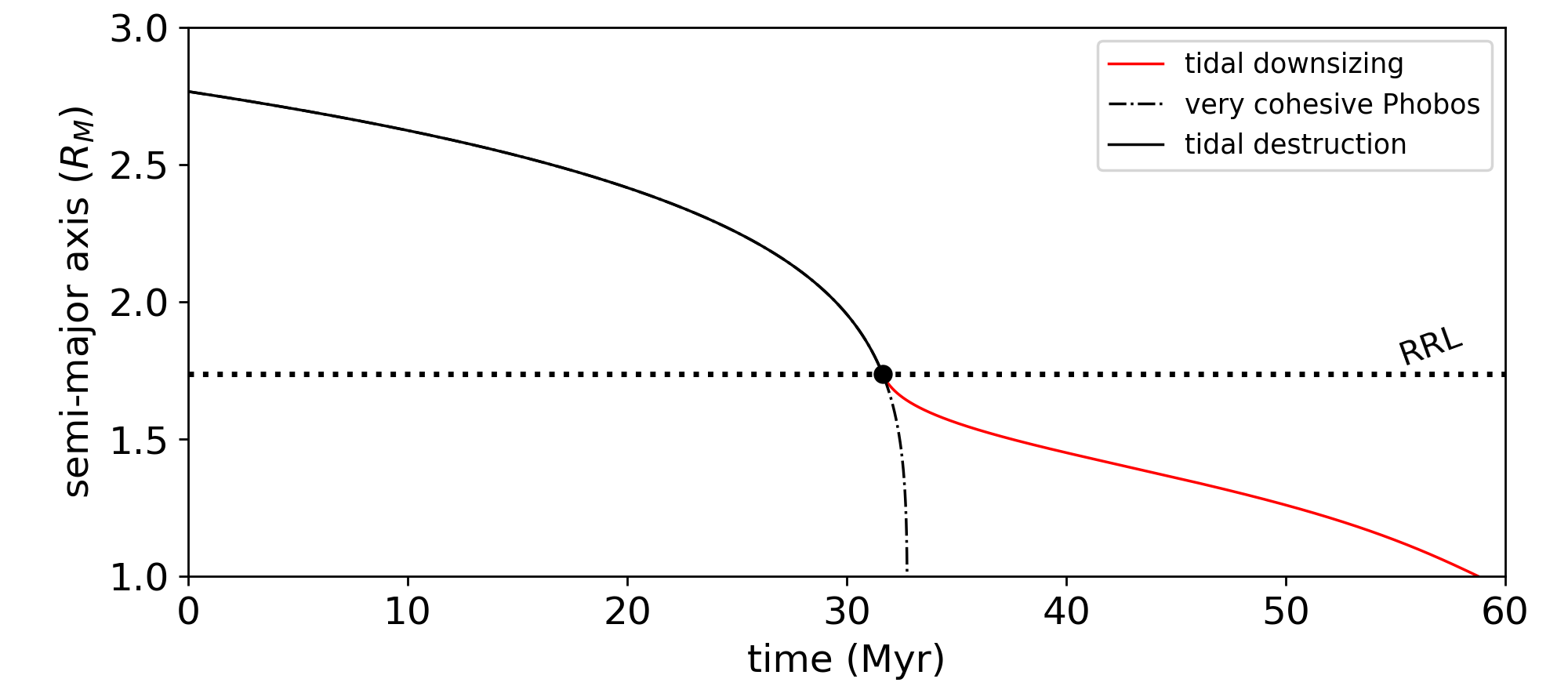}{0.8\textwidth}{(a)}}
\gridline{\fig{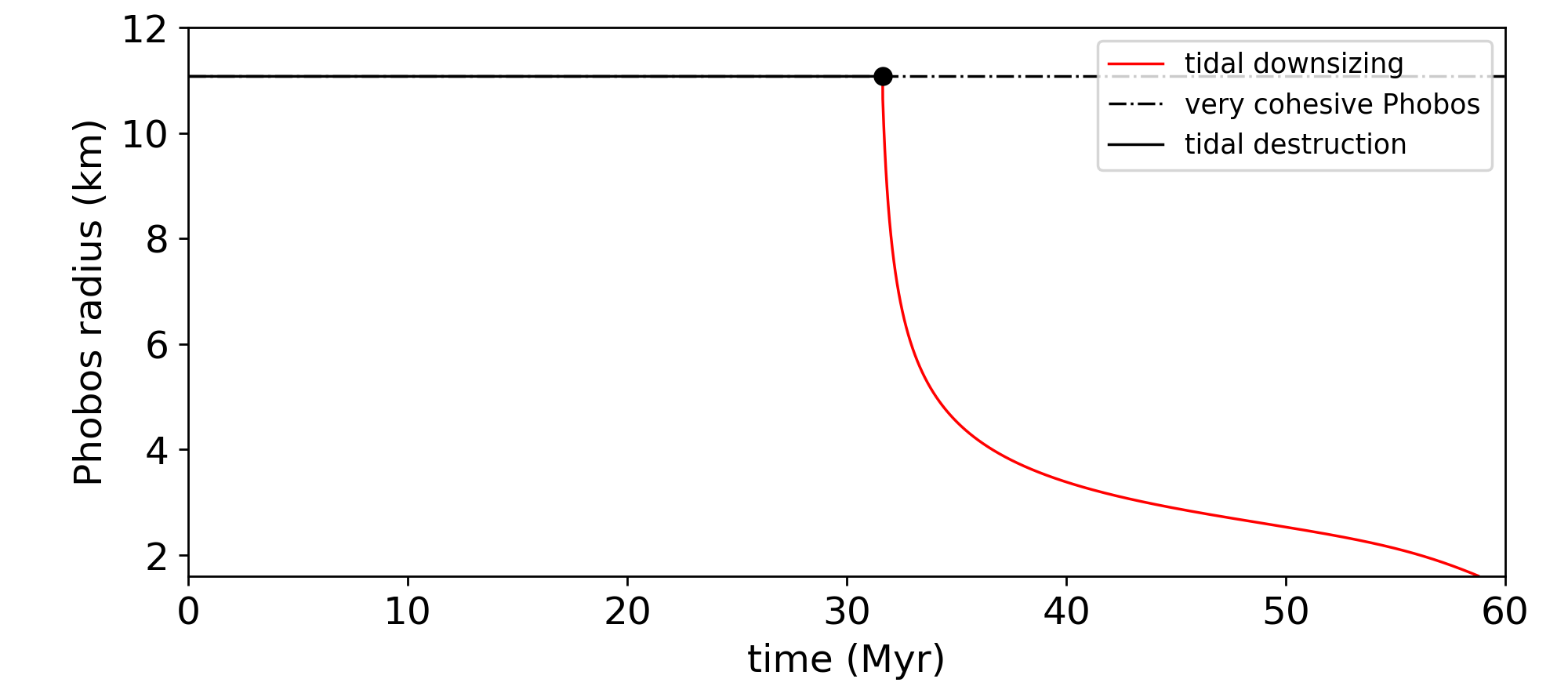}{0.8\textwidth}{(b)}}
\caption{Temporal evolution of (a) semimajor axis and (b) physical radius of Phobos ($k_2/Q=10^{-6}$) under tidal effects. The evolution of the satellite before reaching the RRL (dotted horizontal line) is given by the solid black line, with the black dot showing the instant Phobos would be destroyed. The dash-dotted line shows what the evolution of the satellite would have been if it had not been torn apart by tidal forces. A hypothetical case in which Phobos shrinks due to tidal effects is shown by the solid red line. \label{fig:tidalevolution}}
\end{figure}

Figure~\ref{fig:tidalevolution} shows the tidal evolution of Phobos. We assume $k_2/Q=10^{-6}$ for Phobos \citep{bagheri2021}, where $k_2$ is the Love number and $Q$ is the tidal quality factor. All other quantities are the same as assumed by \cite{bagheri2021}.
The solid black line gives the evolution of Phobos until reaching the RRL ($\phi=40^{\circ}$, dotted line), where it would be completely destroyed according to theory (black point, ``tidal destruction''). The RRL is reached in about 31 Myr. Assuming that the satellite is not destroyed, we show in dashed black line the case where Phobos maintains its original mass \citep[as in][ ``very cohesive Phobos'']{black2015}. In this case, Phobos falls onto the surface in about 32 Myr. Finally, the red line corresponds to the hypothetical case in which Phobos does not disrupt upon reaching the RRL, but only loses its external layers in order to reach a tidal equilibrium shape (''tidal downsizing''). We follow Figure~\ref{fig:rrl} to model the tidal downsizing effect. As Phobos migrates inward, we verify if its mass would allow it to be stable in that location. If not, we assume that Phobos loses the amount of mass necessary for the satellite to be marginally stable. E.g., Phobos with its current mass ($\sim10^{16}$~kg) would not be stable at $\sim 2~{\rm R_M}$ (case with $\phi=25^{\circ}$). Therefore, in our tidal downsizing simulations, we would assume that Phobos eroded to the maximum stable mass at that location, $\sim 6\times10^{15}$~kg. As can be seen, below the RRL, Phobos begins to shrink towards the planet, reaching the surface of Mars at about 60~Myr. When it reaches the planet's surface, its remaining radius is $\sim 2$~km only.

Now, we may wonder what is the size of the fragments produced by tides? Is tidal downsizing a plausible process?

\section{Tidal disruption evolution of a rubble-pile Phobos} \label{sec:pkdgrav}

The tidal disruption location given in Section \ref{sec:rrl} is based on static theories.  As a rubble-pile body disintegrates, the nature of the resulting structural evolution and dynamics of the body and fragments can only be evaluated using numerical modeling. To assess the validity of the tidal downsizing scenario and the size of the fragments produced by tides, we use the \texttt{pkdgrav} $N$-body code to simulate the tidal disruption processes.

Phobos is modelled as a $\sim$11-km-radius self-gravitating rubble pile consisting of $N=18,760$ spherical particles with a $-3$ index power-law size distribution ranging from 220 m to 660 m.  The choice of this size range is based on a compromise between computational cost and model accuracy.  A soft-sphere discrete element model (SSDEM) is applied to compute the contact forces and torques between particles composing Phobos in the normal, tangential, rolling and twisting directions. The material shear and cohesive strengths can be controlled by the given contact parameters. The details of \texttt{pkdgrav} and its SSDEM implementation can be found in \cite{Richardson2000}, \cite{Schwartz2012}, and \cite{Zhang2017, Zhang2018}.

To make direct comparisons with the analyses in previous section, we assign the Phobos rubble pile with a friction angle of 40$^\circ$, and cohesive strength ranging from 1 kPa to 10 kPa.  With the consideration that Phobos' orbit would slightly deviate from a perfect circle during its orbital decay process, we set the orbital eccentricity of the modelled Phobos to 0.0151 (similar to its current orbit) and the orbital semimajor axis to $1.76{\rm R_M}$.  Given that the tidal disruption distance is about $1.74{\rm R_M}$ (see Figure \ref{fig:rrl}), the modelled Phobos is expected to undergo tidal disruption near the periapsis for appropriate cohesive strength.

\begin{figure}[ht!]
\gridline{\fig{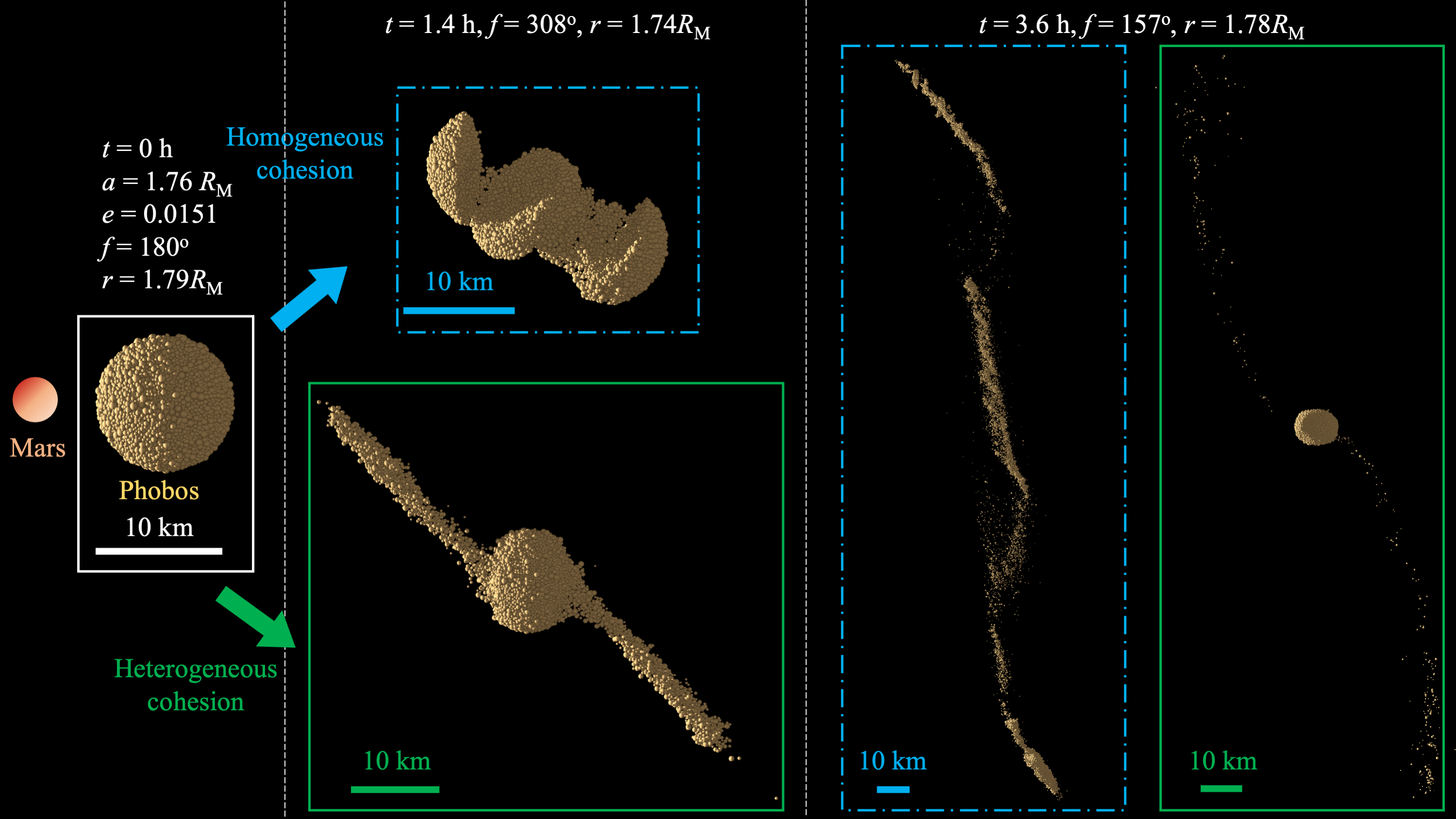}{1.0\textwidth}{}}
\caption{Snapshots of tidal disruption processes for two structures with different cohesion distributions.  Time proceeds from left to right in the three frames separated by the vertical dashed lines.  The time $t$, orbital elements $(a,~e,~f)$, and the corresponding orbital distance are indicated by the white texts on the top of each frame.  The style of the box boundary indicates the initial condition (white solid), the homogeneous cohesion case (blue dotted), and the heterogeneous cohesion case (green solid), respectively. The line of sight is in a direction perpendicular to the orbital plane, and the light rays are ejected from Mars (e.g., in the direction from left to right) to illustrate the Mars' direction. \label{fig:sims}}
\end{figure}

Figure~\ref{fig:sims} presents the structural evolution of a representative case of our tidal encounter simulations when the cohesion is consider to be homogeneously distributed within the rubble pile (indicated by the blue dotted boxes).\footnote{The presented case corresponds to a cohesion value of 10 kPa, which is substantially larger than the assumed cohesion when the RRL is derived, i.e., $\sim$800 Pa, which means the tidal disruption limiting distance predicted by the $N$-body modeling is larger than the theoretical RRL for a given cohesion.  The reason for this quantitative inconsistency could be due to the difference between the continuum and static treatment used in the theoretical derivation of the RRL and the discrete and dynamical treatment used in the numerical model.  The connection between these two treatments is left for future study.}  In response to Mars' tidal forces, the body disrupts internally, and the whole rubble pile disintegrates into a cloud of small fragments that only consist of one or several particles, that is, it disintegrates in its constitutive particles. This is could be due to the fact that the maximum shear stress is located at the body center, and, therefore, cracks tend to initiate internally.  In this case, Phobos could be tidally disrupted entirely when it reaches the RRL. Just as a test, we repeated our simulations, but using an ellipsoidal shape (13.4 x 11.2 x 9.2 km, to resemble the shape of Phobos). We obtain that shape does not affect the tidal disruption behaviors shown in Figure~\ref{fig:sims}.

However, a rubble-pile body that formed through reaccumulation process is unlikely to have a homogeneous structural cohesion distribution.  For example, the geophysical features of the rubble-pile asteroid Bennu observed by the OSIRIS-REx spacecraft indicate that the subsurface cohesion and cohesion in some local internal regions could be much stronger than the surface cohesion \citep{Zhang2022}.  To take into account such a possibility, we assign the rubble-pile Phobos with surface cohesion of 500 Pa and interior cohesion of 30 kPa. The results are shown in Fig.~\ref{fig:sims} (e.g., the green solid boxes). Due to the low strength of the surface regolith and the strong internal structure, the rubble pile only sheds its surface material in response to the tidal forces.  In this case, the tidal disruption process of Phobos would be similar to the one assumed in the tidal downsizing scenario, showing that this is a possible scenario. The shedded materials stay close to the original orbit within 10~h after the tidal disruption, and their follow-up evolution need to be evaluated by the full model. In conclusion, we confirm that the disruption distance of Phobos is qualitatively consistent with the analytical model (see Section~\ref{sec:disruption}). However, in the case of a rubble-pile satellite, N-body simulations show that there is no tidal downsizing: the object fully destroys down to its constitutive particles when it crosses the Roche limit. For tidal downsizing to happen, the object core must be more cohesive than its envelope.

\section{Exploring the recycling model} \label{sec:recycling}
We now turn to the full model, where we include both disk evolution and tidal evolution of the satellites, following an approach similar to \citetalias{hesselbrock2017}. However, whereas only a few tens of Myr evolution could be numerically investigated in \citetalias{hesselbrock2017}, we do perform simulations here on $4.5$~Gyr evolution. Our study is mostly focused on the ring's evolution, which is only little discussed in \citetalias{hesselbrock2017} because of computer limitations.

We perform a set of numerical simulations varying the size of the particles in the debris disk, the friction angle of the rubble-pile (defining the disruption location, i.e., the RRL), and the initial mass of the debris disk. The values assumed by us are given in Table~\ref{cisimul}. The density of disk particles and satellites is assumed to be the bulk density of Phobos ($\rho=1.845~{\rm g/cm^3}$) and the initial disk surface density is defined following \cite{hyodo2017,hyodo2017b}, which is consistent with expectations for a debris disk formed after an impact on Mars forming the Borealis basin, as in \cite{citron2015}.

The simulations are performed using the hybrid code \texttt{HYDRORINGS} \citep{charnoz2010,salmon2010} composed of two self-consistently coupled codes: an one-dimensional finite volume code for tracking the viscous evolution of the disk \citep{salmon2010} and an analytical orbital integrator that follows the satellite evolution \citep{charnoz2011}. The evolution of the disk surface density ($\Sigma$) is calculated on a regular grid composed of 200 uni-dimensional cells \citepalias[same as][]{hesselbrock2017} extending from $1.0~{\rm R_M}$ to $3.2~{\rm R_M}$. The center of a cell has a radial location $R$ while its width is $\Delta R=0.011~{\rm R_M}$. At each time-step, the surface density variation is calculated using a second-order Range-Kutta scheme, using final volume scheme that conserve mass at machine precision. Material falling onto Mars is removed while the material that spreads beyond the FRL (at $3.14~{\rm R_M}$) is converted into one satellite per grid-cell. We verified that total angular momentum is conserved with relative variation smaller than to $10^{-3}$ over 1~Gyr evolution in the simulations presented here.

The temporal variation of the surface density is given by \citep{Bath1981}:
\begin{equation}
\frac{\partial \Sigma}{\partial t}=\frac{3}{R} \frac{\partial}{\partial R}\left[\sqrt{R} \frac{\partial}{\partial R}(\nu\Sigma \sqrt{R})\right]
\end{equation}
where $t$ is time and $\nu$ is the total viscosity. We assume the total viscosity as the sum of the translational, collisional, and gravitational viscosities \citep[see][]{salmon2010}.

Viscosity effects will occur differently if the disk is in the non-self-gravitating or gravitational regime. In the first, the disk is dense enough to be considered a fluid, while in the second, inter-particle interactions become important for the ring evolution. We determine the regime using the Toomre parameter \citep{Toomre1964}:
\begin{equation}
Q=\frac{\Omega\sigma_\nu}{3.36G\Sigma},
\end{equation}
where $\Omega$ is the keplerian frequency and $\sigma_\nu$ is the particle radial velocity dispersion. We have that $\sigma_\nu=2s\Omega$ if $r_h<s$ and $\sigma_\nu=\sqrt{Gm/s}$ if $r_h\geq s$, where $m$ is the mass of the ring particle and $r_h=(2m/3M_M)^{1/3}R$ the particle Hill radius \citep{Daisaka2001}.

For the non-self-gravitating regime ($Q>2$), the total viscosity is \citep{salmon2010}
\begin{equation}
\nu=2.76\frac{\sigma_\nu^2}{\Omega}\left(\frac{s\rho\Sigma}{16s^2\rho^2+9\Sigma^2}\right)+0.75\frac{s\Omega\Sigma}{\rho}
\end{equation}
while for the self-gravitating regime ($Q<2$), it is \citep{salmon2010}
\begin{equation}
\nu=0.81\frac{r_h^5\mathrm{G}^2 \Sigma^2}{s^5\Omega^3}+0.75\frac{s\Omega\Sigma}{\rho}
\end{equation}

We have that the viscous spreading timescale in our system in a crude approximation, will be given by \citep{Brahic1977,salmon2010}:
\begin{equation}
\tau_{\rm vis}=(2.14~R_M)^2\frac{1}{\nu}\sim(2.14~R_M)^2\frac{\zeta\Omega^3}{\Sigma^2} \label{tauvis}
\end{equation}
where $\zeta$ is a function approximately linearly proportional to the particle size $s$.

The semi-major axis $a$ and eccentricity $e$ of the satellites evolve under tidal effects and disk-satellite torques ($\Gamma_s$), and their temporal evolution are given by \citep{Kaula1964,Peale1978,charnoz2010}
\begin{equation}
\frac{da}{dt}=\frac{3k_2M_sG^{1/2}R_M^5}{QM_M^{1/2}a^{11/2}}\left[1+\frac{51e^2}{4}\right]+\frac{2a^{1/2}\Gamma_s}{M_s\left(G M_M\right)^{1/2}}
\end{equation}
\begin{equation}
\frac{de}{dt}=\frac{57k_2\Omega M_sR_M^5}{QM_Ma^5}e+F_{me}
\end{equation}
where $M_s$ is the satellite mass and tidal parameter is $k_2/Q=0.00178$ \citep{bagheri2021}. We use the formalism described in \cite{meyer1987} to estimate the disk-satellite interactions and the toy model described in \cite{charnoz2011} to estimate the eccentricity kicks due to mutual encounters between the satellites ($F_{me}$). Two satellites are considered to merge when the distance between them is less than twice the mutual Hill radius. When immersed in the disk, the satellite accretes material following the recipe given by \cite{thommes2003}. Once the RRL is reached, the total satellite's mass is transferred in the ring-cell in which it is located, and the satellite is removed from the simulation (case without downsizing).

\begin{deluxetable*}{cccc}
\tablenum{1}
\tablecaption{Parameter values assumed in the simulations of Section~\ref{sec:recycling}. ${\rm M_P}$ corresponds to the mass of Phobos, ${\rm M_P}=1.059\times 10^{16}$ \citep{Patzold2014}.  \label{cisimul}}
\tablewidth{0pt}
\tablehead{
\colhead{Parameter} & \colhead{Symbol} & \colhead{Unit}  & \colhead{Values}
}
\startdata
Particle size & $s$ & m & $0.1$, $1$, $10$, and $100$\\
\multirow{2}{*}{Initial disk mass} & \multirow{2}{*}{${\rm M_{disk}}$} & $10^{20}$~kg & $1$, $1.2$, $2$, $3$, and $5$ \\
 &  & $10^{4}~{\rm M_P}$ & $0.9$, $1.1$, $1.9$, $2.8$, and $4.6$ \\
Friction angle & $\phi$ & deg & $25$, $40$, and $80$ \\
\enddata
\end{deluxetable*}

\subsection{Dynamic evolution of an example system}
\begin{figure}[ht!]
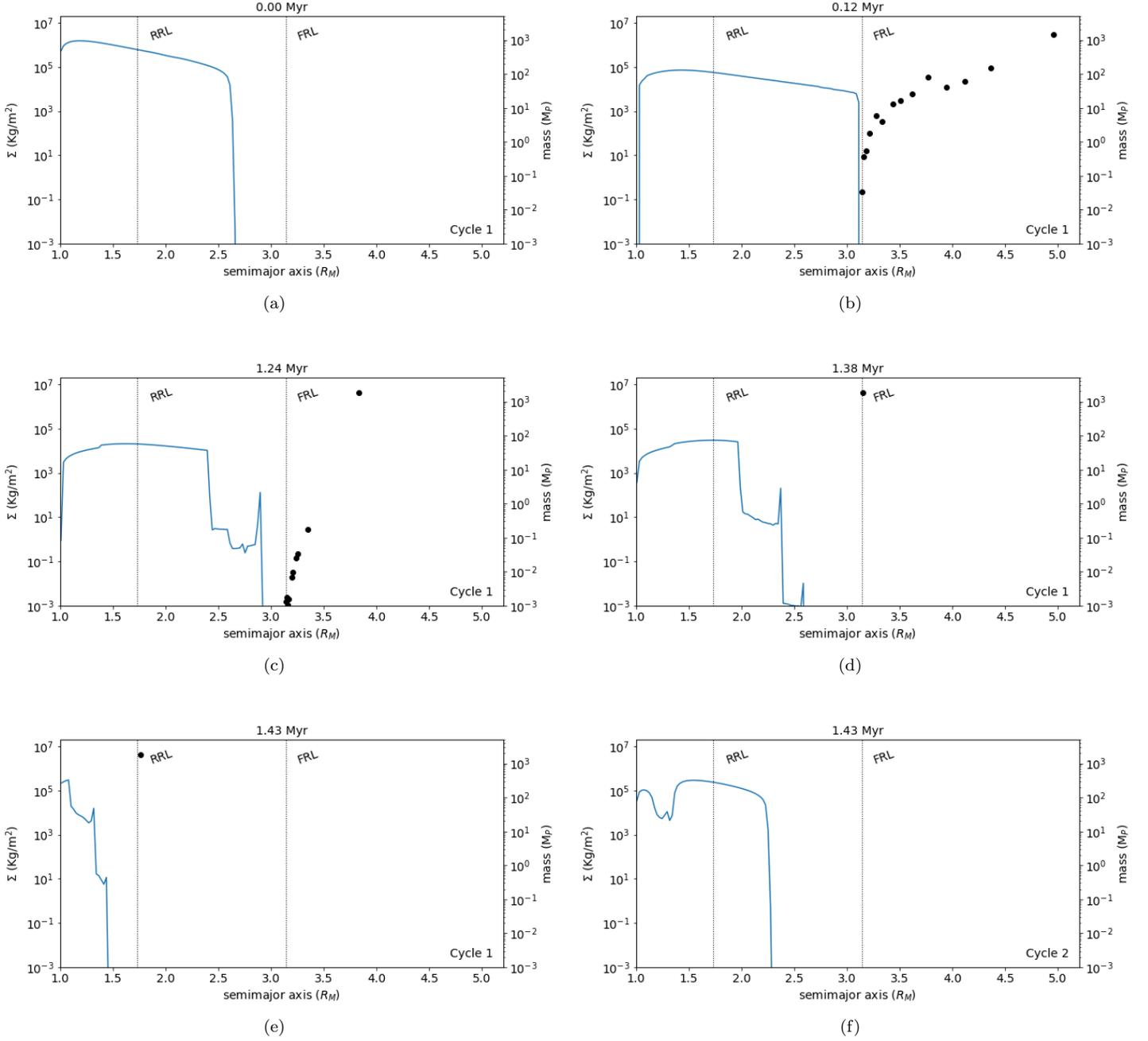

\gridline{\fig{40_10/Zsigma1.png}{0.53\textwidth}{(a)}
\fig{40_10/Zsigma86.png}{0.53\textwidth}{(b)}}
\gridline{\fig{40_10/Zsigma855.png}{0.53\textwidth}{(c)}
\fig{40_10/Zsigma1111.png}{0.53\textwidth}{(d)}}
\gridline{\fig{40_10/Zsigma1197.png}{0.53\textwidth}{(e)}
\fig{40_10/Zsigma1198.png}{0.53\textwidth}{(f)}}
\caption{Evolution of disk surface density (solid blue line, left scale) and satellite mass (black dots, right scale) as a function of distance to Mars (semimajor axis in ${\rm R_M}$). The simulation time is given at the top of each panel, and the vertical dashed lines show the location of RRL ($1.74~{\rm R_M}$) and FRL ($3.14~{\rm R_M}$). The panels only show the first cycle of the simulation, with initial disk mass being $1.1\times 10^4~{\rm M_P}$, particle size $10$~m, and friction angle $40^{\circ}$. An animation with the complete evolution of the system can be found at the link: \url{https://www.ipgp.fr/en/directory/madeira} (30~seconds animation). \label{fig:examplecase}}
\end{figure}

Figure~\ref{fig:examplecase} shows the first cycle (Cycle \#1) of a simulation with ${\rm M_{disk}}=1.1\times 10^4~{\rm M_P}$ \citepalias{hesselbrock2017}, $s=10$~m, and $\phi=40^{\circ}$ (standard model). The left scale of the panels gives the disk surface density ($\Sigma$, solid blue line), while the right scale gives the satellite mass (black dots), as a function of distance to Mars. 

The disk is initially confined within the FRL. Due to its own viscosity, the disk spreads inward and outward (Fig.~\ref{fig:examplecase}a). Some disk material is deposited at Mars equator and some spreads beyond the FRL \citep[e.g.][]{salmon2010}. Beyond the FRL, gravitational instabilities promote rapid accretion of material into aggregates \citep{karjalainen2007,charnoz2010,charnoz2011}, giving rise to Lagrangian moons \citep{rosenblatt2012,hesselbrock2017}. Satellites capture all disk material and neighboring satellites present within their Hill sphere, which depends on both the semi-major axis and planet mass.  The satellite is expected to be porous if the bulk density of disk particles is significantly greater than a critical Hill density defined as $\rho_c=M_M/(1.59a^3)$ \citep{porco2007}. This is our case ($\rho_c=0.334~{\rm g/cm^3}$), which indicates that our assumption of satellites as gravitational aggregates is, in first approximation, applicable.

The moon grows from impacts with other newly formed moons, while migrating outward due to disk-satellite torques. Due to this mechanism, a population of satellites is formed with masses ordered by increasing semi-major axis as can be seen in Fig.~\ref{fig:examplecase}b. Satellite migration slows down as the distance increases, ceasing when the satellite’s 2:1 inner Lindblad resonance (ILR) leaves the disk \citep{charnoz2011}, which happens at ${\rm a_{2:1}^{ILR}}=4.98~{\rm R_M}$. If the disk is sufficiently depleted, the tidal torque exceeds the disk-satellite torque and the satellite migrates inward. Otherwise, it will remain at around $4.98~{\rm R_M}$.

After $\sim0.2$~Myr, the torque balance on the outermost satellite results in inward migration, and the satellite accretes the inner satellites as it migrates toward the planet (Fig.~\ref{fig:examplecase}c). In Fig.~\ref{fig:examplecase}d, the outermost satellite (a Phobos ancestor) remains the only surviving satellite in the system. Here, we define as \emph{Phobos ancestor}, the most massive satellite of the cycle (with a minimal mass of $2~{\rm M_P}$, where ${\rm M_P}$ is Phobos' mass), the one that will migrate inward and reach the RRL. In most cycles, Phobos ancestor is also the outermost satellite at any time. A satellite with these same characteristics, but masses in the range $0.5-2.0~{\rm M_P}$ will correspond to a \emph{Phobos analogue}, according to our definition. This mass range corresponds to the narrowest range needed to ensure that all simulations will produce a satellite analogous to Phobos.

Disk and satellite interact mainly by IRL torques \citep{meyer1987} and if the resonant torque exceeds the disk viscous torque, the satellite confines the disk. The minimum mass for the satellite to confine the disk is given by \citep{Longaretti2018}
\begin{equation}
m_s=1.58\frac{\Sigma}{a_d\rho}\left|\frac{x_e}{a_d}\right|M_M   \label{massedge}
\end{equation}
where $x_e$ and $a_d$ are the distances from the outer disk edge to the satellite and planet, respectively. In the first cycle, a minimum mass of $\sim10^2~{\rm M_P}$ is needed to confine the disk. Such a condition is met by the Phobos ancestor. The $\Sigma$ peaks on Fig.~\ref{fig:examplecase}c-e correspond to the locations of the 2:1, 3:2, and 4:3 ILRs, from left to right. By this mechanism, the satellite pushes the disk towards the planet as it migrates inward (Fig.~\ref{fig:examplecase}e), and material is deposited at Mars equator. The satellite enters in the region below the FRL but is not disrupted before it reaches its RRL ($\sim 1.74~{\rm R_M}$). Just before the satellite reaches its RRL, a residual ring still exists very close to Mars (Fig.~\ref{fig:examplecase}e). Finally, when the satellite crosses its RRL, it is disrupted by tides, and debris are transferred to the ring that spreads rapidly. Then a new cycle begins (Fig.~\ref{fig:examplecase}f). Note that here we have assumed that the satellite is fully destroyed in ring particles when it crosses the RRL, as in the case of the homogeneous cohesion case in Section~\ref{sec:pkdgrav} and \citetalias{hesselbrock2017}. We have also studied the case when the satellite is downsized by tides, but Phobos formation is mostly unchanged (see Appendix~\ref{downsizing}).

\begin{figure}[ht!]
\gridline{\fig{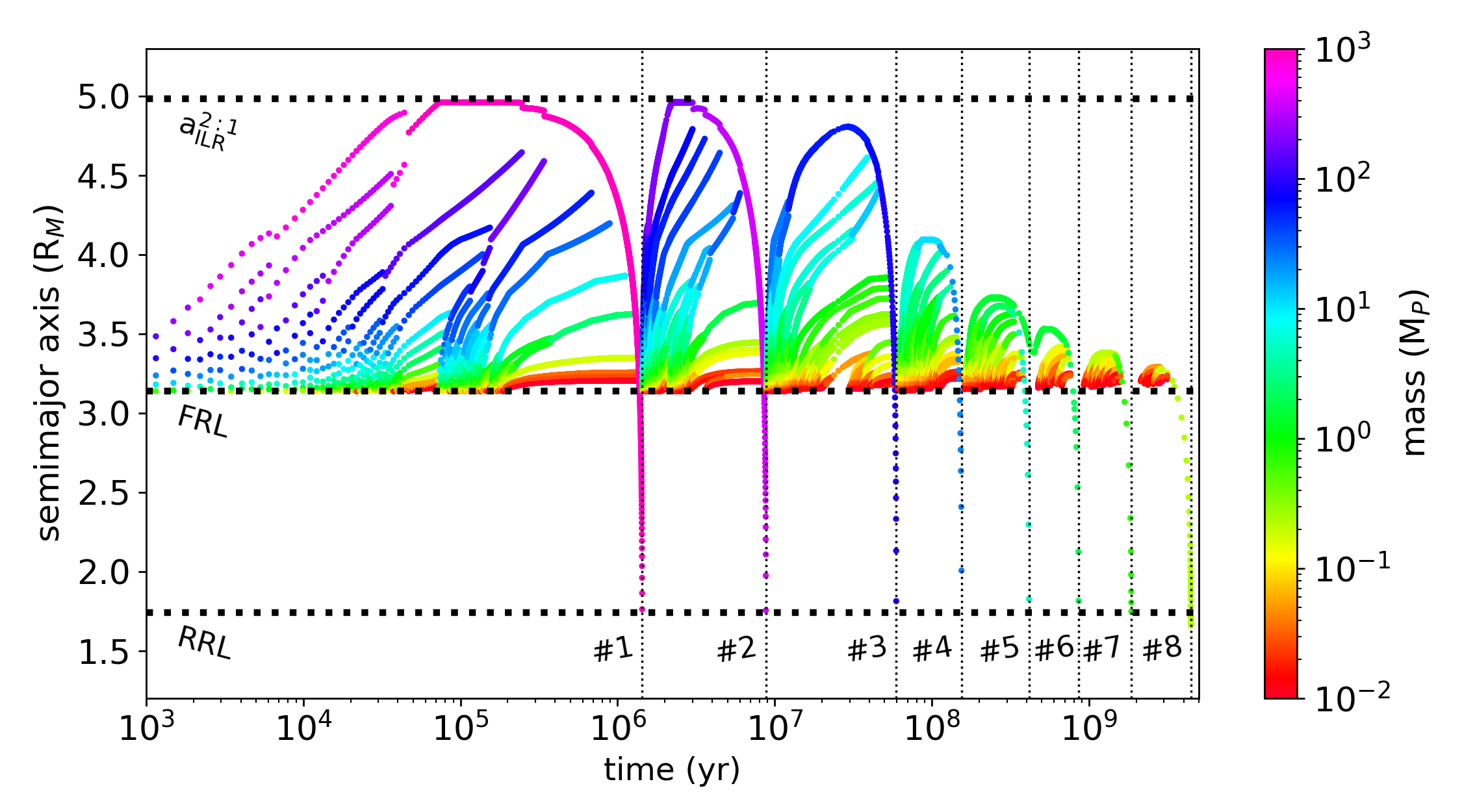}{0.8\textwidth}{}}
\caption{Semi-major axis of satellites as a function of time, for the same simulation as in Figure~\ref{fig:examplecase}. Each dot stands for a satellite obtained in the simulation, but at different times. The color represents the satellite's mass. Satellites with masses similar to Phobos are colored green. The horizontal dotted lines provide the location of RRL, FRL, and 2:1 ILR with FRL. The vertical dotted lines delimit the beginning and end of the cycles. \label{fig:sma40}}
\end{figure}

In general, we find that most cycles follow the same evolution as described above, with some notable exceptions that we  describe hereafter. Figure~\ref{fig:sma40} shows the position and masses of satellites, as a function of time. As we can see, the maximum distance reached by Phobos ancestors decreases with the number of cycles. Indeed, since the disk mass is continuously decreasing, the most massive satellite that can accrete at each new cycle has its maximum mass decreasing with time. This effect was reported in \citetalias{hesselbrock2017}.
\begin{figure}[ht!]
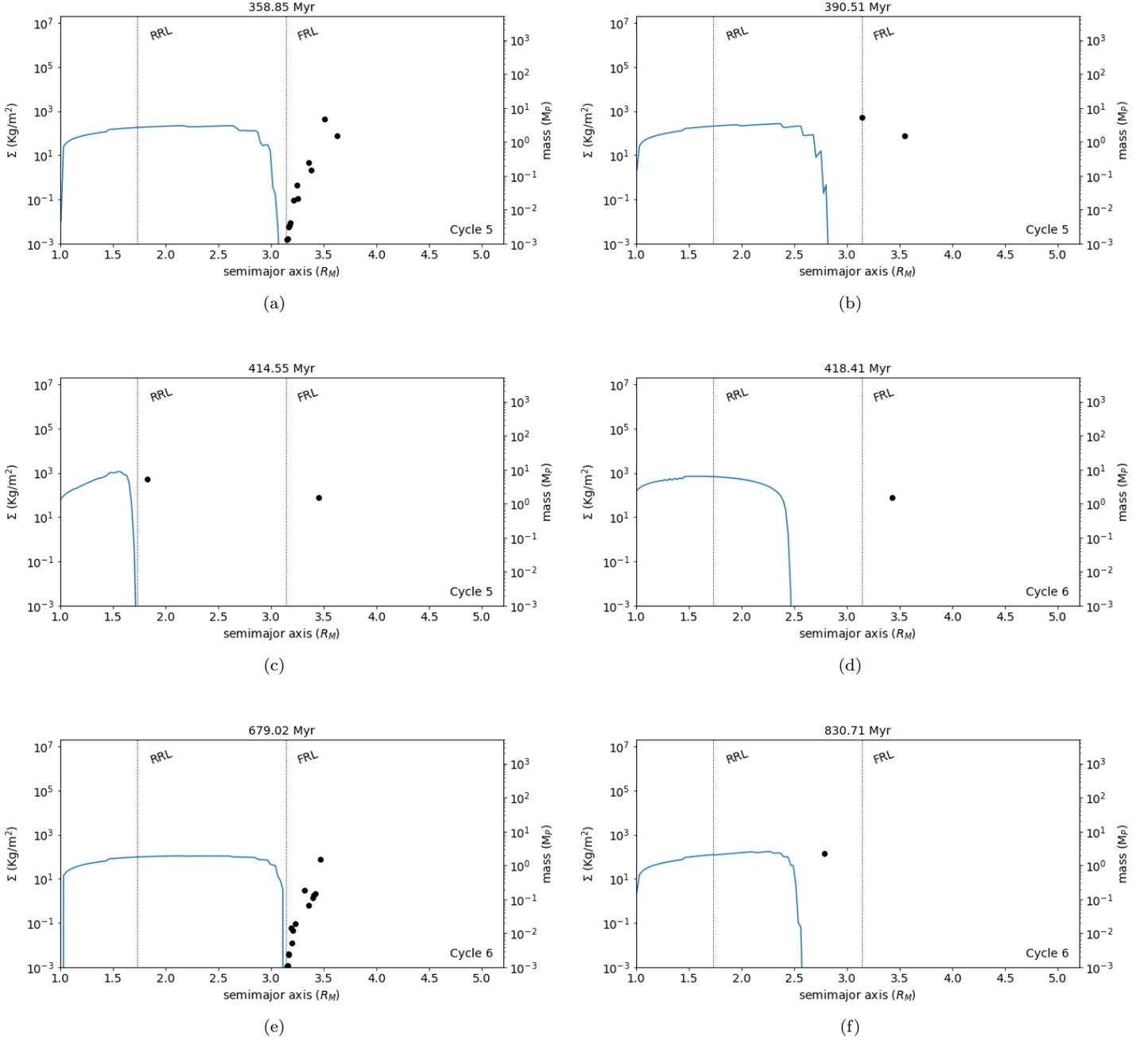

\gridline{\fig{40_10/Zsigma6974.png}{0.53\textwidth}{(a)}
\fig{40_10/Zsigma6979.png}{0.53\textwidth}{(b)}}
\gridline{\fig{40_10/Zsigma6986.png}{0.53\textwidth}{(c)}
\fig{40_10/Zsigma6987.png}{0.53\textwidth}{(d)}}
\gridline{\fig{40_10/Zsigma7574.png}{0.53\textwidth}{(e)}
\fig{40_10/Zsigma7589.png}{0.53\textwidth}{(f)}}
\caption{Evolution of disk surface density and satellite mass, for the fifth and sixth cycles of our standard model. The disk is shown as a solid blue line, with surface density given on the left scale and satellites are given by black dots, with mass on right scale. The vertical dashed lines show the locations of RRL and FRL. An animation with the complete evolution of the system can be found at the link: \url{https://www.ipgp.fr/en/directory/madeira} (30~seconds animation). \label{fig:examplecase_cycle5}}
\end{figure}

In most observed cycles, the disk evolves into a system composed of one Phobos ancestor and a ring confined by the satellite. However, sometimes, a different evolution is observed, like in Cycle \#5 of the standard model.  Figure~\ref{fig:examplecase_cycle5} shows the evolution of disk and satellites in cycles \#5 and \#6, for comparison. Cycle \#6 (Figure~\ref{fig:examplecase_cycle5}d to e), is a ``typical'' cycle, where the most massive satellite is the most distant. In that case, the ring's outer edge is always in 2:1 resonance with the outermost satellite. Cycle \#5 is peculiar because the most massive satellite ($\sim 10~{\rm M_P}$) is not the most distant. The most distant is $\sim 3~{\rm M_P}$ (Figure~\ref{fig:examplecase_cycle5}a and b). Because the most massive satellite feels a stronger planet's tidal torque it moves inward and pushes the ring down to the planet, leaving, temporarily a Roche Zone region almost empty and a satellite with $\sim 3~{\rm M_P}$ just beyond the Roche Limit. If the most massive satellite was very cohesive, it would completely eliminate the ring and would fall onto the planet, just leaving one satellite in the system. However in the current simulation, the object disrupts (Figure~\ref{fig:examplecase_cycle5}c) and Cycle \#6 starts.

Cycle \#5 is very similar to the scenario proposed by \cite{rosenblatt2016}, in which there is only one cycle, because satellites are assumed to be very cohesive, so are not destroyed at the RRL. This is why \cite{rosenblatt2016} find only one remaining Phobos (and one Deimos) and no disk surviving at the end of their process.

In the seventh cycle, a Phobos analogue is obtained at $\sim 1.8$~Gyr after the start of the simulation (assumed to be the time of the giant impact that formed the circum-martian disk). When the satellite accretes the last moon at $\sim 1.8$~Gyr, it is located at $\sim3.2~{\rm R_M}$, taking additional $\sim0.2$~Gyr to reach the actual position of Phobos ($2.76~{\rm R_M}$). Such a value would be the ``age'' of Phobos according to this simulation, which is the time between today and the last major collision with another satellite. Interestingly, this age is close to the lowest age estimated by \cite{ramsley2017} based on craters study. This places the giant impact that hit Mars and formed the Borealis Basin at $t_{\rm gi}\sim2.5$~Gyr after the formation of Mars, while several works suggest that the impact would have happened $<0.5$~Gyr after the formation of the planet (see Section~\ref{sec:discussion}). Phobos' formation must have occurred on a longer timescale than that obtained in the simulation. Furthermore, we get a ring with optical depth of $\tau\sim 9\times 10^{-4}$ coexisting with Phobos, i.e., a ring that would be detectable today. In the following sections, we explore the effects of the particle size, initial disk mass, and friction angle.

\subsection{On the particle size dependence} \label{sec_size}
\begin{figure}[ht!]
\gridline{\fig{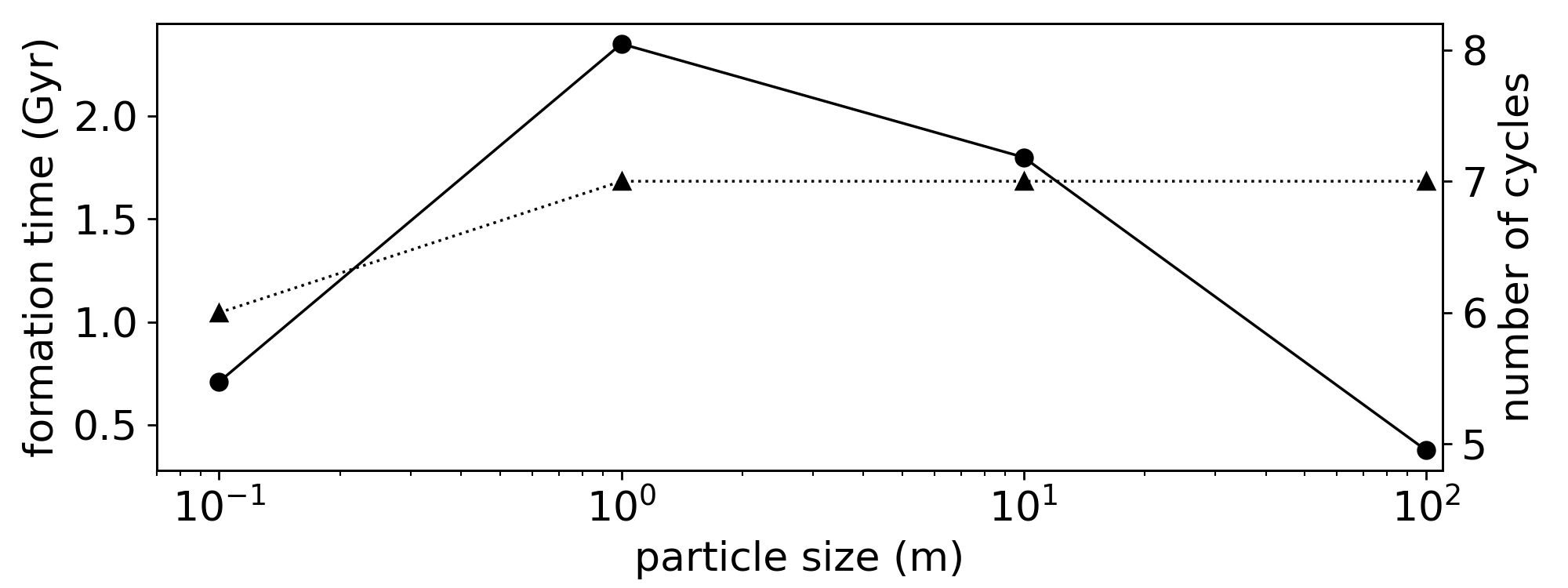}{0.8\textwidth}{(a)}}
\gridline{\fig{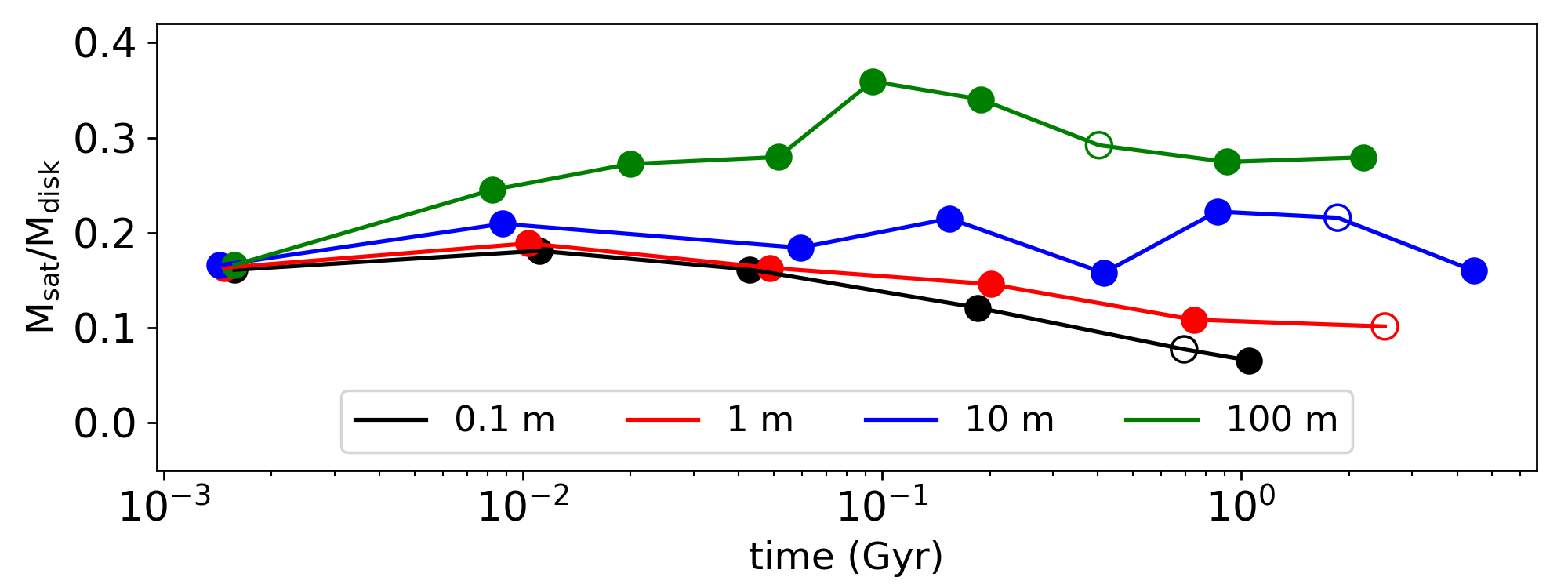}{0.8\textwidth}{(b)}}
\gridline{\fig{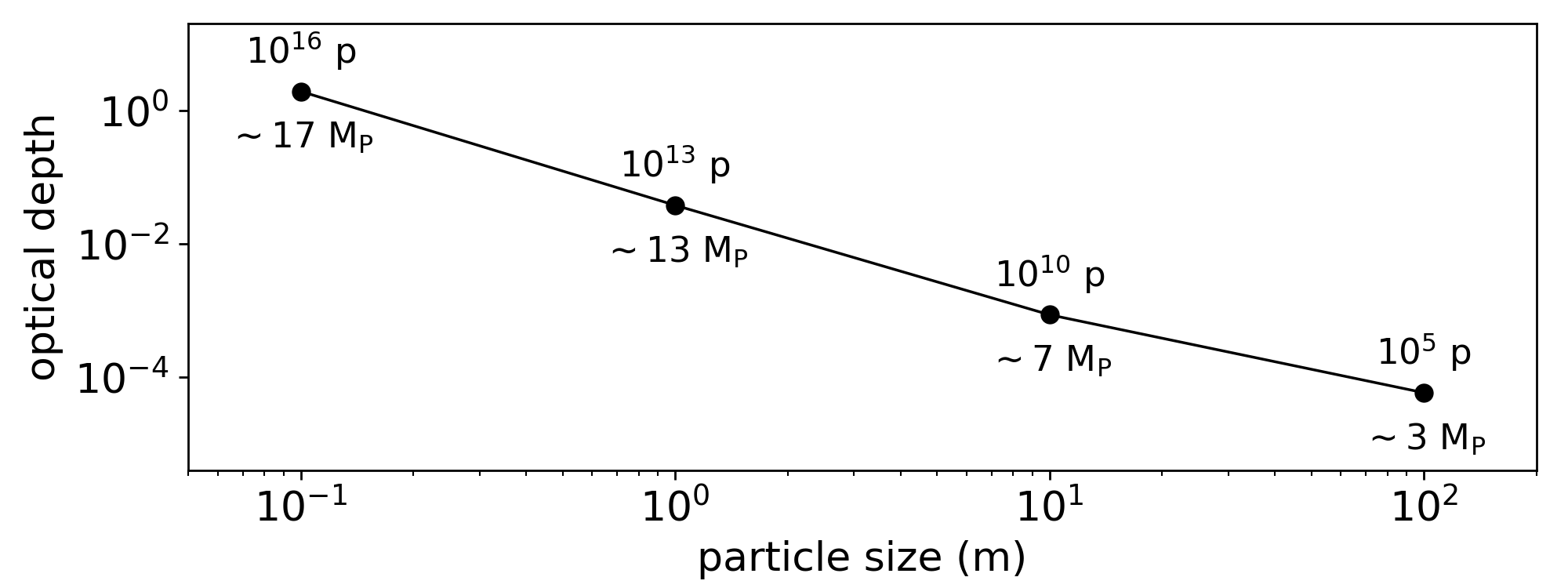}{0.8\textwidth}{(c)}}
\caption{(a) Timespan (left scale, solid line) and number of cycles (right scale, dotted line) to form a Phobos analogue as a function of the particle size. (b) Mass of the largest satellite in the cycle relative to the initial disk mass in the same cycle, with different colors corresponding to different particle sizes. The open points in the panel correspond to the cycles that form a Phobos analogue. (c) Average optical depth of the residual ring coexisting with Phobos as a function of the particle size. We set the initial disk mass as $1.1\times 10^4~{\rm M_P}$ and friction angle equals to $40^{\circ}$. The annotations below and above the black curve give the ring mass and number of particles in the ring, respectively. \label{fig:size}}
\end{figure}

The typical size of the particles in the debris disk is a key parameter of the recycling model, as it defines the viscous evolution of the disk (Equation~\ref{tauvis}). Particles are also the building blocks of the rubble-piles. As shown in Section~\ref{sec:pkdgrav}, a rubble-pile object with homogeneous cohesion breaks down into its constitutive particles, implying that the disk particle size is constant and equal to the assumed initial value. As this parameter is unconstrained, we performed simulations with sizes ranging from $0.1$~m to $100$~m.

Figure~\ref{fig:size} summarizes the results of the simulations with ${\rm M_{disk}}=1.1\times 10^4~{\rm M_P}$ and $\phi=40^{\circ}$. The top panel (Fig.~\ref{fig:size}a) shows on the left scale (solid line) the time to obtain a Phobos analogue, and on the right scale (dotted line), the number of cycles required to form the satellite. The middle panel (Fig.~\ref{fig:size}b) gives the fraction between the mass of the largest satellite formed in one cycle and the initial mass of the disk in the same cycle (${\rm M_{sat}}/{\rm M_{disk} }$), as a function of the time when the largest satellite disrupts. Finally, the bottom panel (Fig.~\ref{fig:size}c) provides data about the ring coexisting with Phobos analogue. The y-axis gives the average optical depth of the ring. Below the black curve is given the mass of the ring, while above it we show the amount of particles that such a mass would represent. Here, we define the average optical depth as the ratio of the total cross-section of material in the ring and the total surface area of the ring, found as
\begin{equation}
\tau=\frac{3M_{\rm ring}}{4\rho sA_{\rm ring}}
\end{equation}
where $M_{\rm ring}$ and $A_{\rm ring}$ are the mass and surface area of the ring extracted from the simulation.

The viscous evolution of the ring through the cycles is a very intricate problem. Although the ring spreads more slowly when the particles are smaller, at first depositing less material onto Mars, we have that for the same reason, satellites will grow more slowly and the cycles will last longer, depositing material on Mars for longer. In addition, the system is influenced by the residual ring obtained at the end of each cycle, affecting interactions with the satellite. Despite this, we obtain some clear relationships between our chosen parameters and the evolution of the system.

In general, we obtain an increase in the Phobos analogue formation time when decreasing the particle size. The exception is the case with $s=0.1$~m, where Phobos analogue is formed faster than the cases with $1$ and $10$~m. This happens because it takes 6 cycles to form Phobos in the case with $s=0.1$~m, while in the other cases, the satellite is formed after 7 cycles. If Phobos were a seventh generation satellite also in the case with $s=0.1$~m, we would get a formation time of $\sim 5$~Gyr, a much longer time than for the case with $s=1$~m. However, the satellite formed in the seventh cycle is too small to be considered a Phobos analogue ($0.4~{\rm M_P}$), also coexisting with a visible ring of $\tau=0.4$. We find that the number of cycles required to form Phobos increases with particle size. That's why the case with $s=0.1$~m requires fewer cycles to form a satellite with mass close to Phobos. In a simulation with $s=1000$~m, for example, we obtain that a Phobos analogue is formed only after 9 cycles.

Our definition of Phobos analogue is somewhat arbitrary and one might wonder whether this could affect our conclusions. The mass range $0.5-2.0~{\rm M_P}$ is the narrowest one required to obtain at least one Phobos analogue in all our simulations. For narrower ranges, some simulations will not form a satellite analogous to Phobos. If we extend the lower/upper limit of the mass range, it takes a longer/shorter period to form Phobos. However, we verify that the main result of the article remains unchanged (Section~\ref{generalremarks}). A visible ring coexisting with Phobos is always obtained, although we obtain a smaller optical depth in cases where Phobos is formed in a longer period (reductions of about 10 times).

In Fig.~\ref{fig:size}b, we can see that ${\rm M_{sat}}/{\rm M_{disk} }$ decreases with time, because of the residual ring that accumulates mass from a cycle to another, reducing the value of the total fraction. If we calculate the ratio between the mass of the largest satellite formed in a cycle and the mass of the largest satellite formed in the previous one, we obtain values $\sim0.25$, the same as those claimed by \citetalias{hesselbrock2017}. This means that the accretion efficiency has a weak dependence on particle size and number of cycles (disk mass). \citetalias{hesselbrock2017} find that the accretion efficiency is mainly affected by the RRL location, a result also obtained by us. The residual ring mass, however, depends on the particle size, which results in the pattern seen in Fig.~\ref{fig:size}b.

Finally, we obtain a clear relationship between particle size and average optical depth. For the same total mass, the cross-section covered by a set of particles is greater for smaller particles. This is, in part, an explanation for our results. Nonetheless, we also obtain an increase in residual ring mass with particle size, explaining the increase in average optical depth with particle size. The simulation with $s=0.1$~m generates a ring with $\tau\sim 2$, of the same order as the denser broad rings of Saturn \citep[rings A and B,][]{French2017}. As we increase the particle size, the average optical depth drops to values of the same order as the C ring, another broad and dense ring of Saturn \citep{Nicholson2014}. For the largest particle size, $\tau=6\times 10^{-5}$ corresponding to a ring composed of $\sim 10^{5}$ particles of $100$ meters of radius. Such material would certainly be detected by orbiters around Mars.

\subsection{On the initial disk mass dependence} \label{sec_mass}
\begin{figure}[ht!]
\gridline{\fig{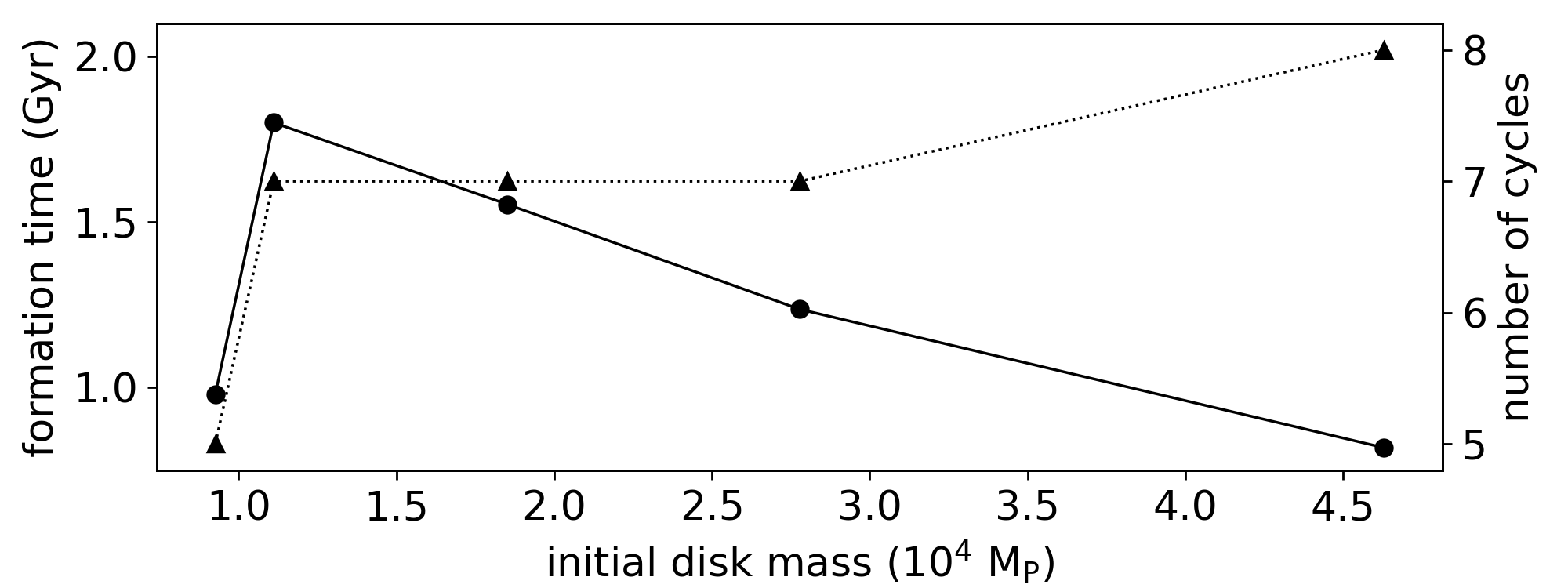}{0.8\textwidth}{(a)}}
\gridline{\fig{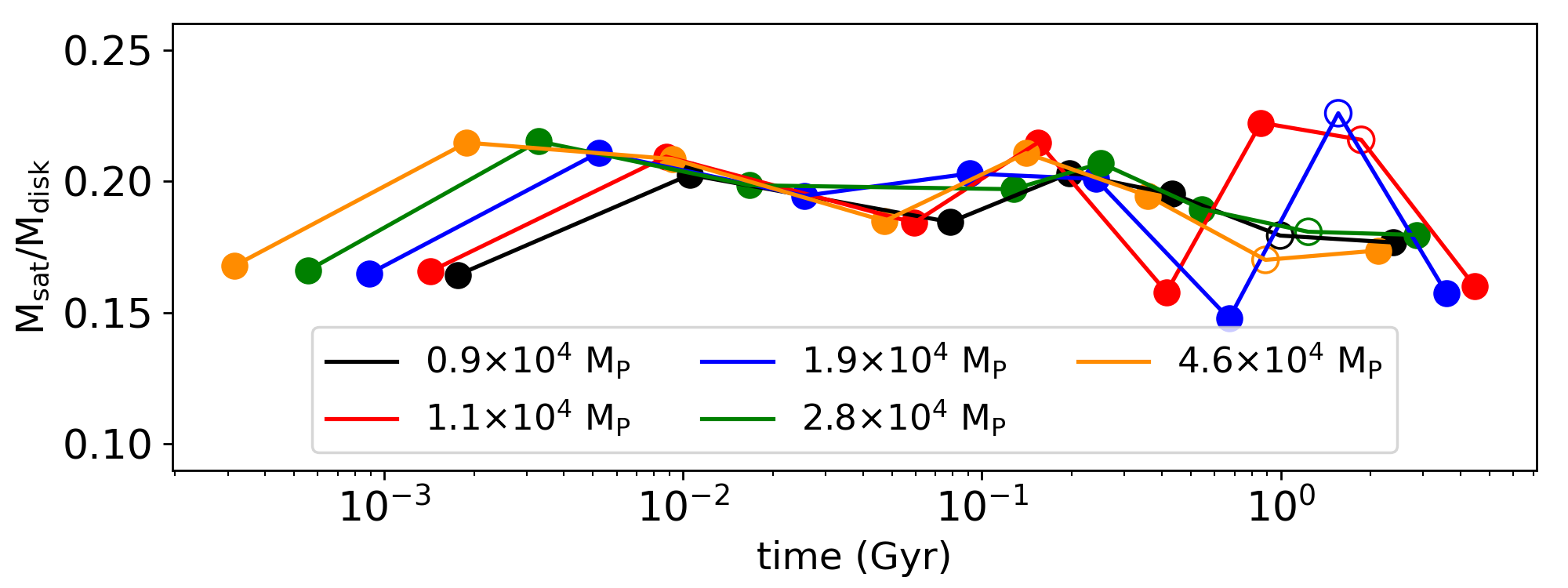}{0.8\textwidth}{(b)}}
\gridline{\fig{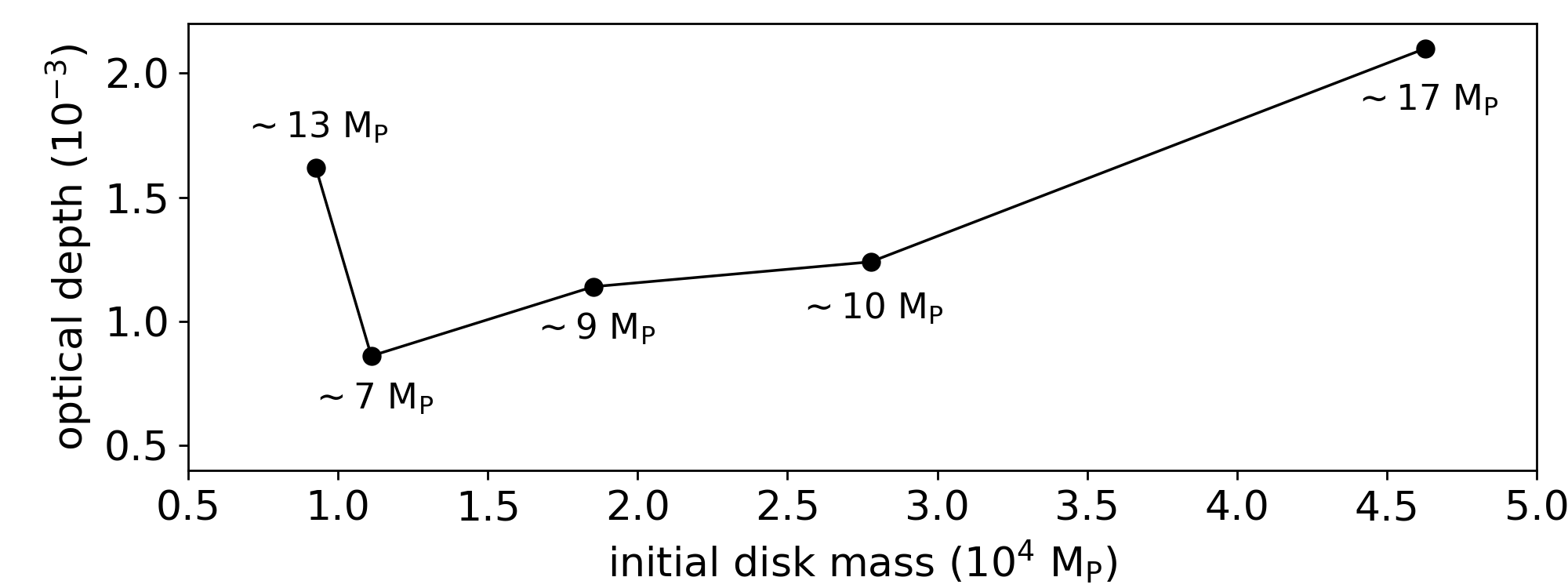}{0.8\textwidth}{(c)}}
\caption{(a) Timespan (left scale, solid line) and number of cycles (right scale, dotted line) to form a Phobos analogue as a function of the initial disk mass. (b) Mass of the largest satellite in the cycle relative to the initial disk mass in the same cycle, with different colors corresponding to different disk mass. (c) Average optical depth of the residual ring coexisting with Phobos. We assumed the particle size as $10$~m and friction angle as $40^{\circ}$. The open points in panel b correspond to the cycles that form a Phobos analogue. The annotations in panel c give the mass of the residual ring.\label{fig:mass}}
\end{figure}

Figure~\ref{fig:mass} shows the same panels as Figure~\ref{fig:size} for simulations with $s=10$~m and $\phi=40^{\circ}$. The greater the mass of the disk, the faster the viscous spreading, being required more cycles to form Phobos. At the same time, we obtain that the formation time decreases with the initial disk mass. Some exceptions are obtained, as in the case with ${\rm M_{disk}}=0.9\times 10^4~{\rm M_P}$ of Fig.~\ref{fig:mass}a. In this case, a Phobos analogue with mass of $1.5~{\rm M_P}$ is obtained in the fifth cycle ($\sim 1$~Gyr). The same cycle in the case with ${\rm M_{disk}}=1.1\times 10^4~{\rm M_P}$ ($\sim 0.4$~Gyr) gives rise to a satellite with a mass of $7~{\rm M_P}$, which is too large to be considered a Phobos analogue, requiring two more cycles to form an object with mass close to that of Phobos. So the formation takes longer in the case with ${\rm M_{disk}}=1.1\times 10^4~{\rm M_P}$.

As can be seen in Fig~\ref{fig:mass}b, there is no clear relationship between ${\rm M_{sat}}/{\rm M_{disk} }$ and ${\rm M_{disk}}$, with all points spreading around an average curve. This leads us to conclude that ${\rm M_{sat}}/{\rm M_{disk} }$ has only a small dependence on the initial disk mass, meaning that the accretion efficiency and the residual ring mass fraction are almost invariant to ${\rm M_{disk}}$. Finally, we get that increasing ${\rm M_{disk}}$, we obtain more massive and brighter residual rings (Fig.~\ref{fig:mass}c). The same exceptions observed for formation time are obtained for the optical depth.

\subsection{On the friction angle dependence} \label{sec_angle}
\begin{figure}[ht!]
\gridline{\fig{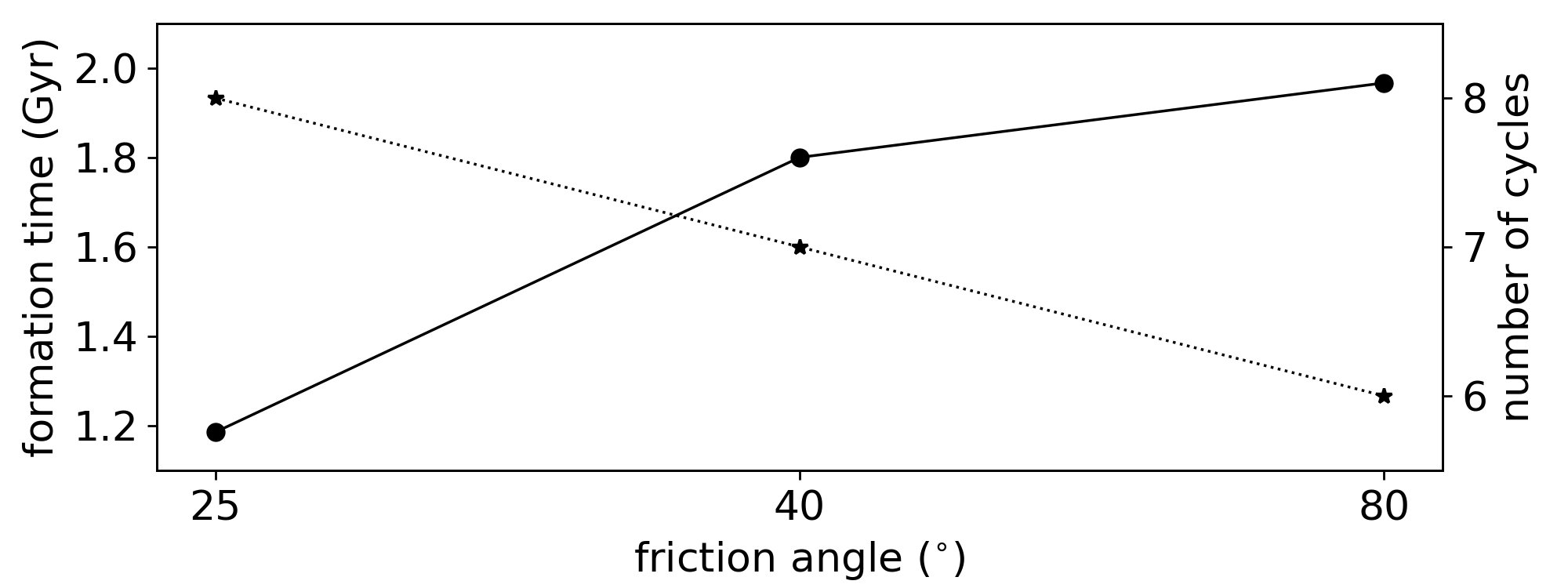}{0.8\textwidth}{(a)}}
\gridline{\fig{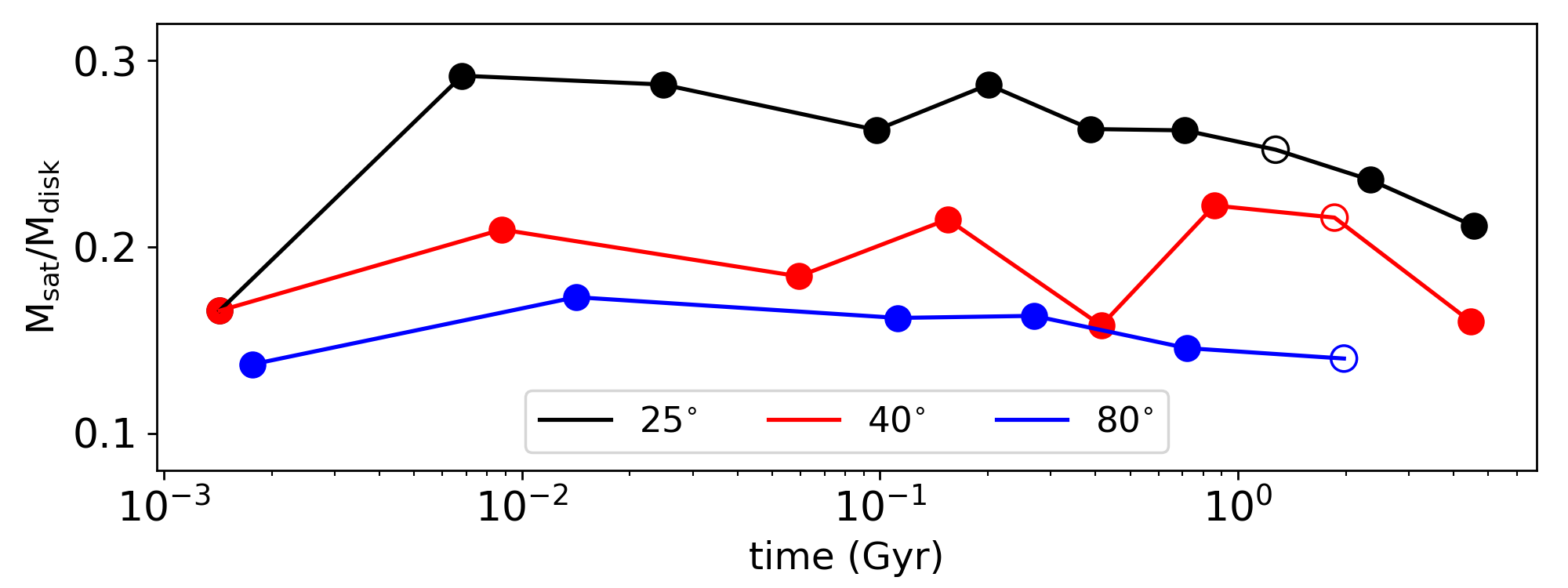}{0.8\textwidth}{(b)}}
\gridline{\fig{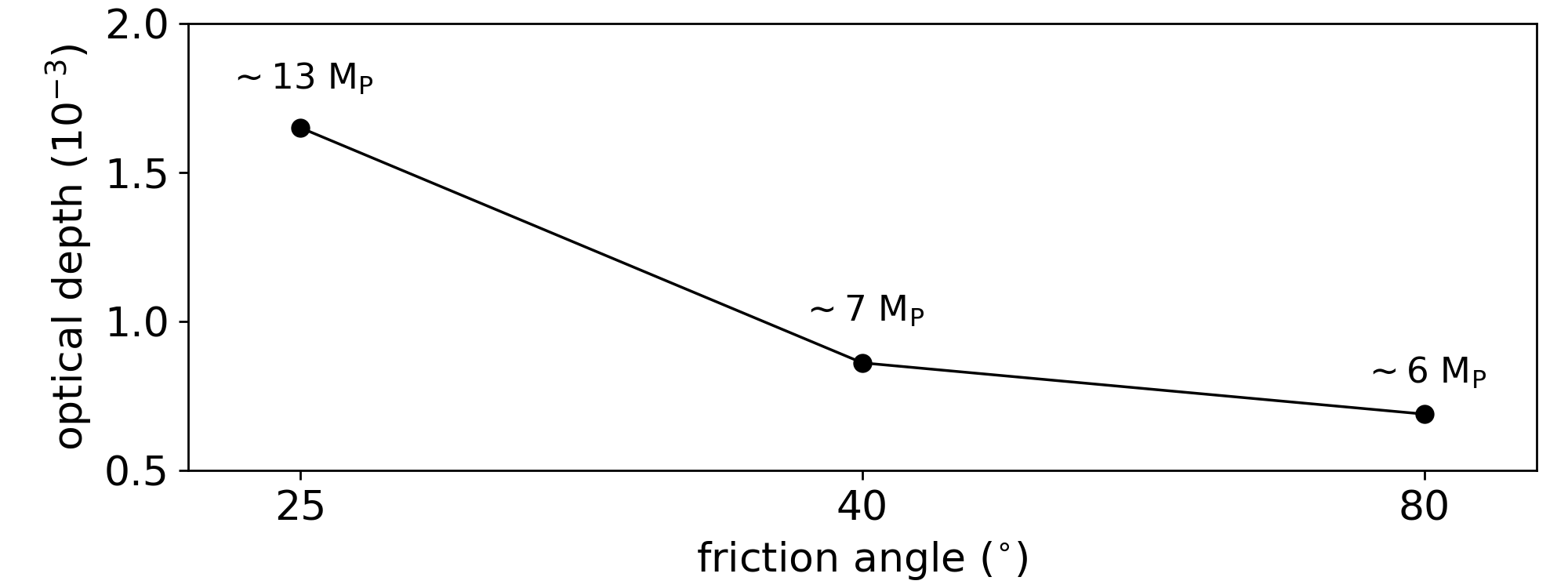}{0.8\textwidth}{(c)}}
\caption{(a) Timespan (left scale, solid line) and number of cycles (right scale, dotted line) to form a Phobos analogue, as a function of the friction angle. (b) Mass of the largest satellite in the cycle in relation to the initial disk mass in the same cycle. (c) Average optical depth of the residual ring coexisting with Phobos. We assumed the initial disk mass as $1.1\times 10^4~{\rm M_P}$ and particle size as $10$~m. The open points in panel b correspond to the cycles that form a Phobos analogue, while the annotations in panel c give the residual ring mass. \label{fig:angle}}
\end{figure}

Now, we analyze the effect of the rubble-pile friction angle on the recycling process. The same panels shown in Figure~\ref{fig:size} are given in Figure~\ref{fig:angle}, for a case with ${\rm M_{disk}}=1.1\times 10^4~{\rm M_P}$ and $s=10$~m varying the friction angle. As discussed in Section~\ref{sec:disruption}, by increasing $\phi$, we are decreasing the location to Mars where the satellite is torn apart by tides. Thus, the closer to Mars the satellite disrupts, the greater the amount of material deposited on the planet per cycle. The more depleted the disk in each cycle, and the slower the viscous spreading. As expected, the number of cycles and ${\rm M_{sat}}/{\rm M_{disk} }$ decrease with decreasing disruption location (Figs.~\ref{fig:angle}a,b), a result also obtained by \citetalias{hesselbrock2017}. In turn, the formation time increases with the friction angle (Figs.~\ref{fig:angle}a), while the optical depth decreases, since the residual ring is less massive when the satellite is closer to the planet (Figs.~\ref{fig:angle}c). In the next section, we summarize the results of our numerical simulations.

\subsection{The ring coexisting with Phobos} \label{ringcoexisting}
In the previous section, we showed that disk accretion on satellites is not completely efficient, leaving a residual ring after the formation of the satellites. In all of our 60 numerical simulations, we obtained a ring coexisting with the Phobos analogues. We then analyzed whether the recycling process would generate an extremely faint ring coexisting with Phobos, which would not be visible with the current observational instruments. Using data from Viking Orbiter 1, \cite{duxbury1988} looked for a ring in the region inside Phobos, ruling out the possibility of a ring with optical depth $\tau>3\times10^{-5}$. More recently, \cite{showalter2006} using Hubble data, did not detect rings coorbital to Phobos and Deimos. They got the upper limit of $3\times10^{-8}$ for a possible Phobos ring. They also ruled out the existence of objects larger than $75$~m in radius around Mars. Such limits define a ``forbidden region'' in terms of optical depth and particle size for a ring around Mars.

\begin{figure}[ht!]
\gridline{\fig{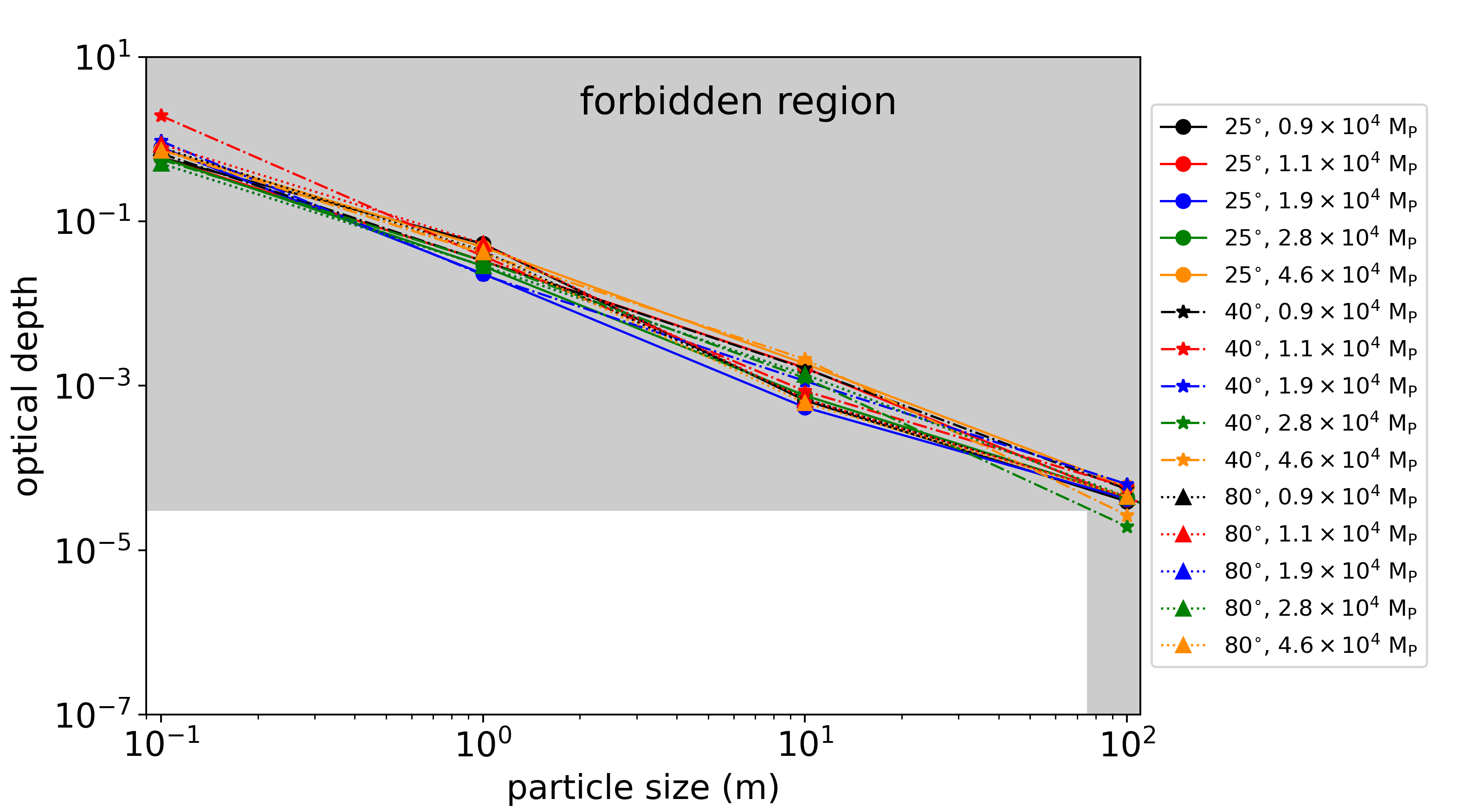}{0.8\textwidth}{(a)}}
\gridline{\fig{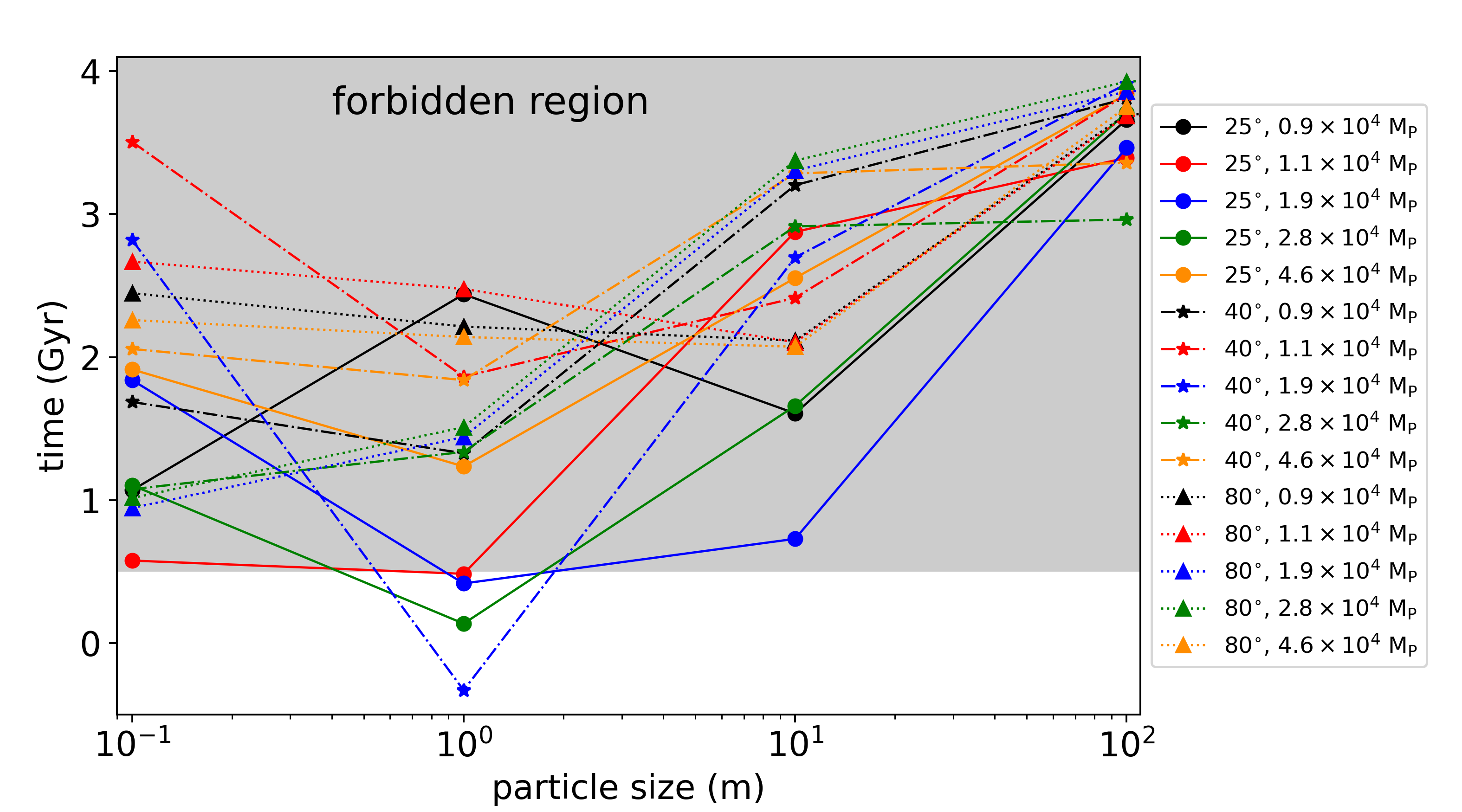}{0.8\textwidth}{(b)}}
\caption{a) Average optical depth of the ring coexisting with Phobos and b) instant of the beginning of the recycling process (after Mars formation), as a function of the particle size. The solid lines with circles, dashed lines with stars, and dotted lines with triangles give the cases with $\phi=25^{\circ}$, $40^{\circ}$, and $80^{\circ}$, respectively. The different colors correspond to different initial disk mass. The gray regions correspond to the forbidden region for the optical depth of the ring and instant of the giant impact. \label{fig:summarize}}
\end{figure}

Figure~\ref{fig:summarize} shows in the top panel the average optical depth of the ring coexisting with Phobos, as a function of the particle size. In the bottom panel, we show the period the recycling process would begin after the formation of Mars. Here, we assume that Mars formed $\sim4.5$~Gyr ago \citep{Nesvorny2018,Izidoro2022}, while we extract from the numerical simulations the time for the formation of Phobos analogue and the time for Phobos to migrate to its current position ($\sim 0.1-0.5$~Gyr). Thus, the instant the recycling process began is given by the relation: $4.5~{\rm Gyr}-{\rm formation~time}-{\rm migration~time}$. The different lines and colors correspond to different sets of disk mass and friction angle, and the gray regions set the forbidden region.
Here, we assume that ideally the recycling process began  $<0.5$~Gyr after the formation of the planet (see Section~\ref{generalremarks}), defining the forbidden region in time (Figure~\ref{fig:summarize}b).

As a rule, we obtain that disks with larger particles ($10$~m and $100$~m) generate a fainter ring and Phobos is formed quickly, while for smaller particles ($0.1$~m and $1$~m), the time for Phobos formation is longer, but the residual ring is brighter. In all our numerical simulations, the optical depth is in the forbidden region. Also, most of our simulations require a very recent giant impact. Only in numerical simulations with $s=1$~m and $\phi=25^{\circ}$, the giant impact would have happened in a time compatible with estimates given in the literature. In the simulation with $s=1$, $\phi=40^{\circ}$, and ${\rm M_{disk}}=1.9\times 10^4~{\rm M_P}$, the Phobos analogue is obtained at its current position in a period greater than the age of Mars. In the next section, we discuss additional processes and implications related to the recycling mechanism.

\section{Discussion} \label{sec:discussion}

\subsection{General Remarks} \label{generalremarks}

We explored the material recycling mechanism for the formation of Phobos, assuming that the satellites aggregated in the debris disk are rubble-piles. In our numerical simulations, the mechanism is successful in forming a satellite with the mass and position of Phobos, however, there are two main caveats: the formation time and the ring coexisting with Phobos.

Regarding the formation time, we need to analyze two important variables: the age of Phobos and the time when the recycling process began. Some studies estimate the age of Phobos through the analysis of topographical structures on the surface of the satellite, but the results differ depending on the method applied. \cite{schmedemann2014}, by counting craters within the Stickney crater -- the largest crater on Phobos' surface -- and assuming that the craters are the result of external impactors, estimate an age of $\sim2.8-4.2$~Gyr for the structure. In turn, \cite{ramsley2017} assume the craters as a result of secondary impacts, estimating an age of $0.1-0.5$~Gyr for the Stickney crater. In our numerical simulations, we obtain that Phobos would have formed $\sim0.1-0.5$~Gyr ago, corroborating the secondary impact crater hypothesis.

The caveat, in fact, resides on the time of the beginning of the recycling process. Works by, e.g., \cite{citron2015,rosenblatt2016} and \cite{canup2018} assume that the debris disk and Borealis basin originated from the same giant impact, which took place $>4$~Gyr ago, according estimates by \cite{nimmo2008,andrews2008} and \cite{marinova2008}. The impacts responsible for the Hellas, Utopia, Isidis, and Argyre basins on Mars \citep{Searls2006,Scheller2020} could also give rise to disks of material, albeit less massive than that from the Borealis impact. In fact, it is likely that the disk that gave rise to Phobos was formed from the collection of these successive impacts. \cite{Bottke2017} using topology analysis, narrow the epoch of the Borealis impact to around $4.5$~Gyr ago, while finding that impacts responsible for the smaller basins occurred between $3.8$ and $4.4$~Gyr ago. Accounting for these studies, we set $0.5$~Gyr after Mars formation as the a maximum fiducial value for the start of the recycling process, which is met only by the cases with $s=1$~m and $\phi=25^{\circ}$.

The other caveat of the recycling model for rubble-pile satellites is the presence of a ring coexisting with Phobos. \cite{duxbury1988} and \cite{showalter2006} set an upper limit for the optical depth of a ring around Mars, but the most likely scenario is that there is not a ring inside Phobos' orbit. This statement is motivated by the presence of orbiters in low-orbits around Mars for long periods of time ($\sim$yrs) -- such as Viking-2 \citep{Christensen1979}, Mars Global Surveyor \citep{Albee2001}, 2001 Mars Odyssey \citep{Mase2005J}, and Mars Reconnaissance Orbiter \citep{Graf2005}. Such orbiters would likely be impacted by debris if there were a ring composed of metric particles around Mars.

Given this, we make a last effort and analyze whether external forces could be responsible for the removal of the ring. Various forces act on rings -- plasma and atmospheric drag, Poynting-Robertson effect, Lorentz force, and Yarkovsky effect \citep{Hamilton1996,Madeira2020,Liu2021,Liang2022} -- however, most are only relevant in the evolution of micrometer-sized particles, a size range not considered by us. This is because rubble-pile satellites are highly unlikely to be reduced to micrometer particles due to tidal forces. For the size range considered by us, the Yarkovsky effect could be the dominant perturbation. We consider this effect in the next section.


\subsection{Debris disk under the Yarkovsky effect}
Yarkovsky effect is a composition of different effects that arise from the asymmetric illumination of particles. The particle side facing a heat source (Sun or Mars) gets hotter than the opposite side, resulting in a thrust in the motion of the particle \citep{rubincam1982}. The strength of the Yarkovsky effect is mainly defined by the linkage of the spin and orbital motion of the particle with the insolation from the Sun or planet. It is the strongest only when the particle rotates with undisturbed principal axis rotation \citep{Bottke2002}. Considering a disk of material, we have that collisions are responsible for the tumbling of the particles, which weakens the Yarkovsky effect that can become insignificant if the tumbling timescale is smaller than the orbital period \citep{rubincam2014}.

To assess whether the Yarkovsky effect is significant for the debris disk, a complete study of the spin variation of the particles would be necessary, which is beyond the scope of this work. However, for completeness, we study the case in which the tumbling of particles due to collisions is disregarded. The spin vector of the particles is assumed to be constant and perpendicular to the equatorial plane during the simulation. Although physically inconsistent -- collisions between particles are what induces the disk viscous spreading -- we can consider that the actual evolution of the disk is likely to be a scenario between this case and the one given in Section~\ref{sec:recycling}.

We have redone all the simulations of Section~\ref{sec:recycling} including the Yarkovsky effect. For this, we computed the secular variation of the semi-major axis due to two different components: the Yarkovsky–Schach and the seasonal Yarkovsky. The Yarkovsky–Schach effect is related to Solar illumination, where one side of the particle absorbs sunlight and remits it in the infrared when in the planetary shadow, feeling a kick that increases its semi-major axis. The seasonal Yarkovsky effect results from the remission by the particle of photons from Mars illumination, this effect being responsible for the orbital decay.

The semimajor axis variation is \citep{Rubincam2006}
\begin{equation}
\frac{da}{dt}=\frac{F_{\rm sun}B\sin\delta}{18nc\rho s}\left[2(1-A_v)b_1\sin\theta_{\rm sun}-\left(1.78+\frac{2.06A_M}{1-A_M}\right)\left(\frac{R_M}{a}\right)^2(1-A_M)(1-A_{IR})\right] \label{yarko}
\end{equation}
where $b_1$, $B$, $\theta_{\rm sun}$, and $\delta$ are functions presented in \cite{Rubincam2006}. The parameters assumed in our calculations are given in Table~\ref{parameters}.
\begin{deluxetable*}{cccc}
\tablenum{2}
\tablecaption{Assumed parameters for the Yarkovsky effect. \label{parameters}}
\tablewidth{0pt}
\tablehead{
\colhead{Parameter} & \colhead{Symbol} & \colhead{Value}  & \colhead{Reference}
}
\startdata
Solar insolation on Mars & $F_{\rm sun}$ & $590.3$~Wm$^{-2}$ & \cite{Rubincam2006} \\
Mars Bond albedo & $A_M$ & $0.248$ & \cite{delgenio2019} \\
Average temperature & $T_0$ & $229$~K & \cite{rubincam2014} \\
Specific heat & $C_p$ & $690.8$~Jkg$^{-1}$K$^{-1}$ & \cite{rubincam2014} \\
Thermal conductivity & $K$ & $2.54$~Wm$^{-1}$K$^{-1}$ & \cite{rubincam2014} \\
Infrared emissivity & $\epsilon_{IR}$ & $0.9$ &  \cite{rubincam2014} \\
Visible albedo & $A_v$ & $0.05$ & \cite{rubincam2014} \\
Infrared albedo & $A_{IR}$ & $0.1$ & \cite{rubincam2014}
\enddata
\end{deluxetable*}

We obtain that the Yarkovsky effect is responsible for the decay of the orbits for the set of parameters assumed by us. For the largest particle sizes ($s=10$~m and $100$~m), the decay timescale due to the Yarkovsky effect $\tau_Y\sim10^8$~yr. Such a value is greater than the viscous spreading timescale of the first cycles ($1-5$), which, therefore, show the same evolution obtained in Section~\ref{sec:recycling}. The timescale $\tau_Y$ is comparable to the viscous spreading timescale only for the cycle forming Phobos and the previous one. In these cycles, the amount of material that falls on Mars increases, reducing the average optical depth of the ring. The average optical depth reduction is $<30\%$, and the ring coexisting with Phobos still remains in the forbidden region for all cases with $s=10$~m and $100$~m.

The picture is different for the cases with the smallest particles. $\tau_Y\sim10^{6}$~yr and $\tau_Y\sim10^{7}$~yr for particles of $0.1$~m and $1$~m, respectively. It implies that the Yarkovsky effect is relevant already in the third cycle. As the Yarkovsky effect removes material, the disk spreads more slowly, forming less massive satellites. Therefore, the cycles become slower, resulting in a more efficient action of the Yarkovsky effect, giving rise to a ripple effect. In the end, we get that the disk material is completely removed after 3-5 cycles.

In general, the results of our simulations with $s=0.1$~m and $1$~m fall into three different cases:
\begin{itemize}
    \item  In the cases with RRL closer to the planet ($\phi=40^{\circ}$ and $80^{\circ}$) and less massive disk (${\rm M_{disk}}=0.9-2.8\times10^{4}~{\rm M_P}$), we find that the disk is quickly removed ($\sim$50~Myr), with the last satellite formed being too massive to be considered a Phobos analogue. That is, the disk is completely removed before forming Phobos.
    \item In the cases with $\phi=40^{\circ}$ and $80^{\circ}$ and more massive disk (${\rm M_{disk}}=2.8-4.6\times10^{4}~{\rm M_P}$) and in the cases with $\phi=25^{\circ}$ and ${\rm M_{disk}}=0.9-1.9\times10^{4}~{\rm M_P}$, the disk is completely removed in $\sim10^8-10^9$~Gyr and a Phobos analogue is obtained without the ring. That is, the disk is completely removed in the cycle that forms Phobos.
    \item For cases with $\phi=25^{\circ}$ and ${\rm M_{disk}}=1.9-4.6\times10^{4}~{\rm M_P}$, the disk is also completely depleted in $\sim10^8-10^9$~Gyr, but the last satellite formed is too small to be considered a Phobos analogue. The disk is completely removed in a cycle after the one that forms Phobos.
\end{itemize}
\begin{figure}[ht!]
\gridline{\fig{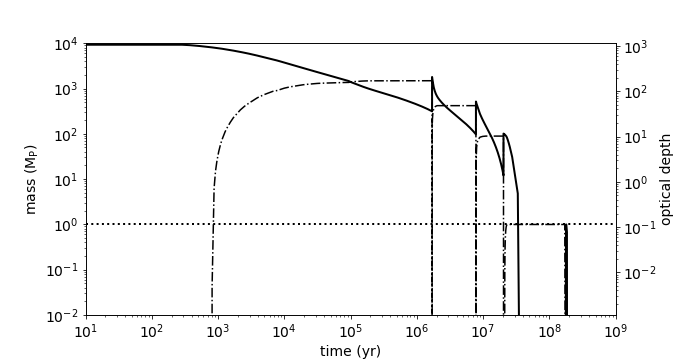}{0.8\textwidth}{}}
\caption{Evolution of disk and satellite masses as a function of time for a simulation with ${\rm M_{disk}}=0.9\times 10^4~{\rm M_P}$, $\phi=25^{\circ}$, and $s=0.1$~m, including the Yarkovsky effect. The left scale gives the mass, with the solid line corresponding to the mass of the disk and the dashed line to the mass of the satellites. The horizontal dotted line shows the mass of Phobos. The right scale gives the optical depth of the disk. An animation with the complete evolution of the system can be found at the link: \url{https://www.ipgp.fr/en/directory/madeira} (17~seconds animation). \label{fig:dissipation}}
\end{figure}

Figure~\ref{fig:dissipation} shows on the left scale the disk mass (solid line) and the satellite mass (dashed line), for a system with ${\rm M_{disk}}=0.9\times 10^4~{\rm M_P}$, $\phi=25^{\circ}$, and $s=0.1$~m. The right scale gives the optical depth of the disk, and the horizontal dotted line places the mass of Phobos. We find that the evolution of the system in the first two cycles is almost the same as in the case without the Yarkovsky effect, which is due to the fact that the cycle timespan is shorter than $\tau_Y$. However, for the third cycle, the timescales of viscous spreading and the Yarkovsky effect timescales are comparable, and we can see a more abrupt drop in the disk mass curve. In this case, we obtain a less massive Phobos ancestor than the one formed in the case without dissipation.

In the fourth cycle, $\tau_Y$ exceeds the spreading timescale and the disk is completely removed in $\sim46$~Myr. When it occurs, there are 15 satellites with radial mass ranking. In the absence of the disk, satellites only migrate due to tidal effects, with the farthest one (at $3.3$~R$_M$) migrating faster due to its greater mass. It accretes the internal ones. At $90$~Myr, the Phobos analogue is formed (at $3.2$~R$_M$) with a mass of $M_{p}$, reaching the current Phobos location at $\sim150$~Myr. The satellite is destroyed in $\sim178$~Myr, giving rise to a disk that is completely removed in a few Myr.

In some simulations with Yarkovsky effect, Phobos is formed without a ring, but in these cases the satellite would be very young. For example, in the case of Figure~\ref{fig:dissipation}, we get a Phobos that is only $60$~Myr old. In fact, it is the presence of the disk in the simulations of Section~\ref{sec:recycling} that delays satellite migration due to tides, allowing ages in the range constrained by \cite{ramsley2017}. The simulation of Figure~\ref{fig:dissipation} also requires the recycling process beginning $150$~Myr ago.

As already discussed, we include the Yarkovsky effect assuming the extreme (and unrealistic) case where tumbling due to collisions can be disregarded. In the real case, collisions will generate tumbling of the particle spin, damping the Yarkovsky effect. Depending on the impact configuration, collisions can also result in fragmentation and grinding into smaller dust and the latter must be affected by external forces that do not play a role in the evolution of metric particles. For example, \cite{Liang2022} study the effect of solar radiation and Poynting-Robertson force on the evolution of particles orbiting Mars, finding that particles with $s\lesssim 100~\mu m$ have lifetimes of up to $\sim10^4$~yr. Therefore, such effects can act over the cycles, removing material resulting from impacts between particles.

It seems general that by removing material from the system, Phobos can be formed without a ring, but in a short timespan, regardless of the external effect included. However, the viscous evolution of the debris disk over the cycles is an intricate problem and we do not rule out the possibility that a set of external forces combine a reasonable formation time for Phobos with the lack of a ring. A more appropriate study of the particulative evolution of the ring with Yarkovsky effects, including particle tumbling, collisional grinding, and other external forces is needed to verify this possibility.

\subsection{Resonances with Deimos}
Models that assume Phobos formation beyond the 2:1 MMR with Deimos \citep{craddock2011,rosenblatt2016,canup2018} obtain that Phobos would have gone through such a resonance $\sim2$~Gyr ago. Due to this, Deimos' eccentricity would be increased to values of $\sim0.002$ \citep{yoder1982}, requiring an intense dissipation in the satellite, in order to dampen the eccentricity to its current value \citep[$2.7\times 10^{-4}$,][]{jacobson2014}. In turn, \citetalias{hesselbrock2017} obtain that only the oldest  Phobos ancestor crosses the 2:1 MMR with Deimos ($>4$~Gyr ago), which diminishes the value of dissipation factor $k_2/Q$ required for Deimos.

In our simulations, because of the ring that remains in every cycle, Phobos ancestors migrate further comparing to the results of \citetalias{hesselbrock2017}. As a consequence, the 2:1 MMR crossing no only happens in the first cycle, but also in the second and third -- as can be seen in Figure~\ref{fig:examplecase}. To analyze the effect of the resonance crossing, we performed some N-body numerical simulations with the Rebound code \citep{rein2012}, using the MERCURIUS hybrid symplectic integrator. We include Mars, Deimos with its current semi-major axis, a Phobos ancestor initially at $4$~R$_M$, tidal effect, and an artificial force to mimic the disk-satellite effect \citep[also see][]{cuk2020}. Both satellites are initially in near circular and near equatorial orbits.

\begin{figure}[ht!]
\gridline{\fig{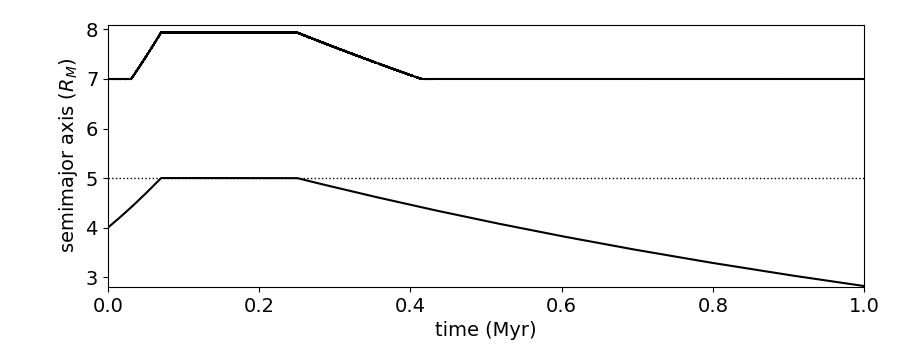}{0.8\textwidth}{(a)}}
\gridline{\fig{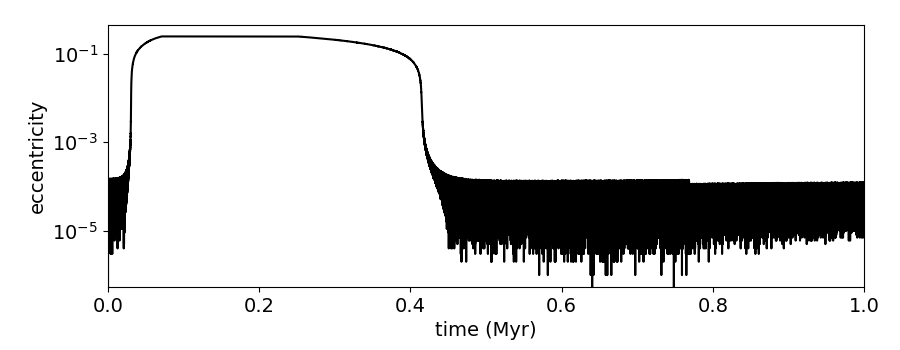}{0.8\textwidth}{(b)}}
\gridline{\fig{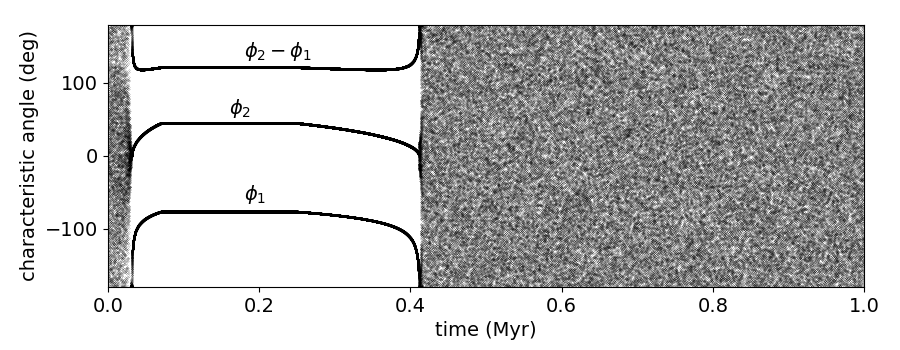}{0.8\textwidth}{(c)}}
\caption{Temporal evolution of a) semimajor axis of the Phobos ancestor and Deimos, b) eccentricity of Deimos, and c) characteristic angles associated with the 2:1 MMR (see text). Phobos ancestor is initially at $4$~R$_M$ and Deimos is at $\sim 7$~R$_M$. The N-body simulation approximately reproduces part of the cycle~1 shown in Fig.~\ref{fig:examplecase}. \label{fig:mmr}}
\end{figure}

Figure~\ref{fig:mmr} shows the case of a system with an artificial force responsible for approximately reproducing the evolution of Phobos ancestor in the first cycle of Figure~\ref{fig:examplecase}. The top panel (Fig.~\ref{fig:mmr}a) shows the temporal evolution of the semi-major axis of Phobos ancestor (at $4$~R$_M$) and Deimos (at $\sim7$~R$_M$), the middle panel (Fig.~\ref{fig:mmr}b) shows Deimos' eccentricity and the bottom panel (Fig.~\ref{fig:mmr}c) shows the following characteristic angles: $\phi_1=2\lambda_D-\lambda_P-\varpi_P$, $\phi_2=2\lambda_D-\lambda_P-\varpi_D$, and $\phi_2-\phi_1=\varpi_D-\varpi_P$. $\lambda$ and $\varpi$ are the mean longitude and the argument of pericenter, respectively, while the subscripts $_P$ and $_D$ refer to Phobos and Deimos, respectively.

Phobos ancestor crosses the 2:1 MMR location ($\sim4.4$~R$_M$) after $\sim0.03$~Myr, capturing Deimos in an apsidal corotation resonance. This resonance occurs when both characteristic angles $\phi_1$ and $\phi_2$ librate, meaning that the satellite periapses are fixed relative to each other \citep[$\phi_2-\phi_1$,][]{ferrazmello1993}. When the inner migrating satellite is orders of magnitude more massive than the outer one -- which is the case of all our simulations -- capture will result in an asymmetric apsidal corotation, which means that the argument of pericenter of the satellites will not remain aligned or anti-aligned relative to each other \citep{ferrazmello2003}.

While migrating outward, the Phobos ancestor pushes Deimos outward and the eccentricity of both satellites increases \citep{Beauge1994,Beauge2006}, as can be seen in Fig.~\ref{fig:mmr}. Upon reaching $\sim5$~${\rm R_M}$, in $\sim0.07$~Myr, Phobos migration ceases and so does the growth of eccentricities, with the eccentricity of Deimos oscillating around $0.245$ while Phobos' eccentricity reaches values up to  $3\times 10^{-5}$. At this point, Deimos is located at $\sim7.9$~R$_M$. Asymmetric corotation is stable only for outward migration and eccentricities start to decrease when Phobos starts to migrate inward after $0.25$~Myr. $\phi_2-\phi_1$ begins to precess, and the resonance is broken when Deimos is again at $7$~R$_M$, having eccentricity $\sim 10^{-4}$.

Analog dynamics is verified for the second and third cycle. At the end of the latter, Deimos is approximately restored to its initial position, while its eccentricity has been increased to $\sim 10^{-3}$. We find that for $k_2/Q=1.2\times10^{-4}$, Deimos' eccentricity reaches a value compatible with the observed one. Such a value of $k_2/Q$ corresponds to that estimated by \cite{bagheri2021} for a loosely connected aggregate Deimos.

Similar results were obtained in all simulations, leading us to conclude that the recycling model is compatible with the hypothesis that Deimos is a direct fragment of the impact. \cite{rosenblatt2016} assume that the impact gave rise to a population of embryos beyond the fluid Roche limit and that a massive innermost satellite migrating outward captured the embryos at 2:1 MMR, forming Deimos. We emphasize that in the recycling model, this may also be a possibility, with a Phobos ancestor acting as the innermost satellite proposed by \cite{rosenblatt2016}.

\subsection{Limitation of the code}

The \texttt{HYDRORINGS} code is extensively described in \cite{salmon2010} and in  \cite{charnoz2011} and seems very similar to the code used in \citetalias{hesselbrock2017}. The ring is described using an hydrodynamical approach \citep{salmon2010}, adapted to compute long term evolution, on timescales comparable to viscous spreading timescales. Particles in rings are assumed to have a single size in order to make the calculation tractable, and also because a hydrodynamic formalism with multiple sizes still does not exist (in a closed-form). The main limitations concern the orbital  evolution of satellites. Satellites' semi-major axes evolve under the combined influence of planet's tides, and disk's torque (summed over all first order mean motion resonances implanted in the rings). Satellites mutual interactions are not considered. Including theses interactions would imply to integrate their motion over billions of years, representing 10 to 100 billions of orbits, which is beyond current computer capacities (if the disk must be tracked simultaneously). These limitations will be addressed in the future, but still, would imply significant theoretical and numerical developments. Therefore, the results presented in this paper should be considered as a first order study, like any previous study of this kind. It's not possible to conserve total mass and total angular momentum at the same time in our code due to the cell grid approach. Because of this, we use an adaptive step size that controls the conservation of angular momentum between two successive steps. In the simulations presented here, we obtain that the total angular momentum is conserved with a relative variation smaller than $10^{-3}$ over 1~Gyr evolution. To ensure the statistical validity of our results, we ran a large number of simulations and on different computers. Our tests show that the results are not affected by the resolution adopted in the simulation, demonstrating the robustness of the numerical results.

\section{Conclusion} \label{sec:conclusion}
In this work, we have analyzed the material recycling model for the formation of Phobos, initially proposed by \cite{hesselbrock2017}. We focused our study on the evolution of the debris disk and we have assumed  that the parent moons of Phobos  are  rubble piles. Due to tidal forces, rubble-pile satellites are ground down to their constitutive particles (or quickly tidally downsized, depending on their cohesion and constitutive characteristics). We have explored the effect of particle size and friction angle onto the evolution of the debris rings and satellite evolution, as well as the disk initial mass. As described in \cite{hesselbrock2017, hesselbrock2019} we do find that an ancient moon, heavier than Phobos, would experience multi-cycle recycling process when it crosses the Roche Limit. At each cycle, a new ring and a new satellite population is formed at the Roche limit. At every cycle the ring and moon mass diminishes consistently with \cite{hesselbrock2017}.

Our main result is that the disk that is produced after the tidal destruction of the parent bodies of Phobos, never fully converts into satellites nor falls onto Mars. In every cycle, a remaining Roche-interior ring orbits around the planet, co-existing with one or several moonlets just exterior to the Fluid Roche Limit. By comparing our simulations to observational surveys of dust or ring around Mars \citep{duxbury1988, showalter2006}, we show that the debris ring resulting from the recycling process should have been already detected, if the recycling process did really happen. Indeed, when Phobos is formed, the resulting ring is either too bright or either its constituent particles are too big to be reconciled with observations. Whereas we varied many parameters of the simulation (particle size, angle of friction, initial mass), we never find a case where the final ring could be reconciled with observations.

This  raises the question : why do we not see a ring around Mars today? One could argue that the Roche-interior debris ring could have been removed after the formation of Phobos. However, in the recycling model, Phobos is formed between 3.0 and 3.5 Mars radii, leading to tidal infall time $<0.5$~Gyr, and no known process seems able to remove macroscopic particles (cm to m) on such a timescale. The only solution we found is to consider the Yarkovsky effect acting on macroscopic particles. However, it would be only effective if the rotation state of particles does not change on Gyr timescales, which doesn't seem possible in a collisional system. Also, it would require the recycling process beginning a few Gyr after the formation of Mars, while there are no evidence of a giant impact in the recent history of Mars.

Therefore, we conclude that Phobos is unlikely to be the result of a recycling process. The impacts that gave rise to the basins seen on Mars surface certainly produced a Roche-interior disk of material in the few hundred thousand years after Mars formation \citep{andrews2008,marinova2008}. This disk may be responsible for the formation of satellites including Phobos, but Phobos is unlikely to have experienced recycling as in \cite{hesselbrock2017}. Phobos would be older than predicted by the recycling model, perhaps in the range obtained by \cite{schmedemann2014}. Also, it would have formed far beyond the Fluid Roche Limit ($\sim3.2~{\rm R_M}$).

If the satellites formed from the Roche-interior disk were rubble-pile objects, the recycling mechanism would be expected to have happened, with the disk diminishing in mass at each cycle until its (almost) disappearance. Note that it is consistent with our results when we do not constrain an object with a mass similar to Phobos to be the real Phobos. Now, if the formed satellites were very cohesive, they would push the disk towards Mars and then fall entirely onto the planet \citep[as in][]{rosenblatt2016}. So the recycling mechanism would not happen at all. Such a scenario is very similar to the stirred disk model and could be an explanation for the elongated craters seen on Mars.

Martian Moons eXploration (MMX), developed by the Japan Aerospace Exploration Agency (JAXA), is expected to be launched in 2024. The MMX mission plans to collect samples of $>10$ g from the surface of Phobos and return them to Earth in 2029 with the aims of elucidating the origin of Martian moons \citep{fujimoto2019}, collecting geochemical information about the evolution of Martian surface environment \citep{hyodo2019}, and searching for traces of Martian life \citep{hyodo2021}. The MMX data will be decisive in constraining the physical properties of Phobos, allowing the distinction between the stirred disk and the recycling models.

\begin{acknowledgments}
G.M. thanks FAPESP for financial support via grants 2018/23568-6 and 2021/07181-7. R.H. acknowledges the financial support of MEXT/JSPS KAKENHI (Grant Number JP22K14091). R.H. also acknowledges JAXA's International Top Young program. P.M. acknowledges funding support from CNES and from the European Union's Horizon 2020 research and innovation program under grant agreement No. 870377 (project NEO-MAPP). S.G.W. thanks FAPESP (2016/24561-0) and CNPq (313043/2020-5). Thanks to the Brazilian Federal Agency for Support and Evaluation of Graduate Education (CAPES), in the scope of the Program CAPES-PrInt, process number 88887.310463/2018-00, International Cooperation Project number 3266.  Numerical computations were partly performed on the S-CAPAD/DANTE platform, IPGP, France.  The \texttt{pkdgrav} simulations were performed on M{\'e}socentre SIGAMM hosted at the Observatoire de la C{\^o}te d'Azur. Visualization of Fig.~\ref{fig:sims} was produced using the POV-Ray ray-tracing packing. We thank Julien Salmon for his review that increased the quality of this paper.
\end{acknowledgments}

\vspace{5mm}

\software{\texttt{HYDRORINGS} \citep{charnoz2010,salmon2010}, \texttt{pkdgrav} \citep{Richardson2000,Schwartz2012,Zhang2017,Zhang2018}, \texttt{Rebound} \citep{rein2012}}



\appendix
\section{Effect of tidal downsizing on Phobos formation} \label{downsizing}
In Section~\ref{sec:disruption} we theorize the possibility that the tidal force is responsible for satellite downsizing inside the RRL, in a process we called ''tidal downsizing''. In fact, we show in Section~\ref{sec:pkdgrav} through soft-sphere numerical simulations that this is possible depending on the cohesion and internal structure of those satellites. So, we also performed numerical simulations accounting for the tidal downsizing effect.

Upon reaching the RRL, we assume that the satellite loses the amount of material necessary to be marginally stable at that location (Fig.~\ref{fig:rrl}), and the mass is transferred to the ring-cell in which it is located. Then, at each time-step, the satellite mass is changed according to Fig.~\ref{fig:rrl} and the eroded mass is transferred to the ring.
\begin{figure}[ht!]
\gridline{\fig{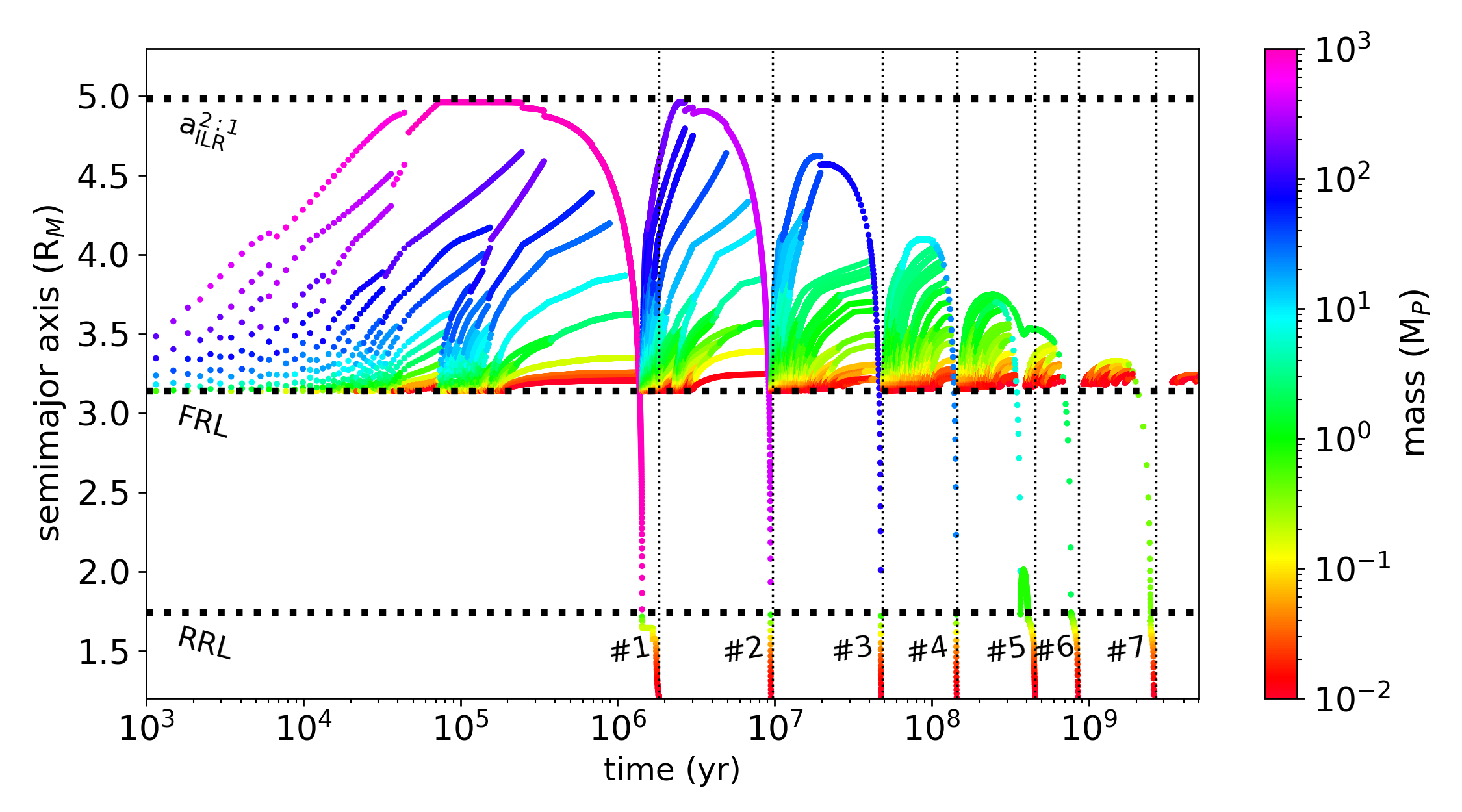}{0.8\textwidth}{}}
\caption{Semi-major axis of satellites as a function of time, for a simulation with ${\rm M_{disk}}=1.1\times 10^4~{\rm M_P}$, $s=10$~m, and $\phi=40^{\circ}$, with tidal downsizing. Each dot stands for a satellite obtained in the simulation, but at different times. The color represents the satellite's mass. An animation with the complete evolution of the system can be found at the link: \url{https://www.ipgp.fr/en/directory/madeira} (28~seconds animation).\label{fig:downsizing}}
\end{figure}

Figure~\ref{fig:downsizing} shows a simulation with the same parameters as our standard model (Fig.~\ref{fig:sma40}), including the tidal downsizing effect. Generally speaking, we find that the downsizing has a minor effect on Phobos formation and evolution, as can be seen by comparing Figs.~\ref{fig:downsizing} and \ref{fig:sma40}. The maximum mass that can survive at the RRL ($1.74~{\rm R_M}$) is about $10^{16}$~kg, a value at least two orders of magnitude smaller than the mass of Phobos ancestor in the first four cycles. In these cycles, Phobos ancestor loses more than $99\%$ of its mass right at the RRL, which is very close to the extreme hypothesis of full destruction in Fig.~\ref{fig:sma40}.

Small differences in the system can be noticed in the cycles prior to Phobos formation -- cycles \#5 and \#6 in Fig.~\ref{fig:downsizing}. The differences, however, are not enough to significantly change the results regarding the formation time and optical depth (Section~\ref{ringcoexisting}). In these cycles, more than 10\% of the mass of the ancestor satellite survives when reaching RRL, but due to the rapid decay of mass as a function of the semi-major axis in Fig.~\ref{fig:rrl}, we have that the satellites lose more than 99\% of their mass before reaching $\sim1.6~{\rm R_M}$. Physically, this means that tidal effects are responsible for rapid downsizing of the body when inside the RRL.

It is also seen that the surviving parts of the satellites are not massive enough to confine the ring, being immersed in it. The satellites don't promote a cleaning process by pushing material towards the planet, as one might think. Although tidal downsizing does not have a significant effect on the formation of Phobos, it might be important in the post-evolution of the satellite, as already pointed out in Section~\ref{sec:disruption}.

\bibliography{refs}{}
\bibliographystyle{aasjournal}



\end{document}